\documentclass[lettersize,journal]{IEEEtran}
\usepackage{amsmath,amsfonts, bm}
\usepackage{array}
\usepackage{listings}
\usepackage{adjustbox}
\usepackage[group-digits=true,group-minimum-digits=4]{siunitx}
\usepackage{tikz,pgfplots}
\usepackage{textcomp}
\usepackage{stfloats}
\usepackage{url}
\usepackage{verbatim}

\usepackage{graphicx} 
\usepackage{float}
\usepackage{subcaption}

\usepackage{acronym}
\usepackage{caption}
\usepackage{comment}
\usepackage{soul}
\usepackage{color}
\usepackage{mathtools}
\usepackage{balance}
\usepackage{multicol}
\usepackage{longtable}
\usepackage{amssymb}
\usepackage{booktabs, makecell, tabularx}
\usepackage{longtable}
\usepackage[section]{placeins}
\usepackage{nomencl}
\usepackage{footmisc}
\usepackage{cite}
\usepackage{xcolor}
\usepackage{caption}
\usepackage{setspace}
\usepackage{amsthm}
\usepackage{academicons}
\usepackage[acronym]{glossaries}
\usepackage{mathtools, cuted}
\usepackage{lipsum}
\usepackage{balance}
\usepackage[font={footnotesize},labelfont={footnotesize}]{caption}
\usepackage{xpatch}
\usepackage[colorlinks,
            urlcolor=blue,
            linkcolor=blue,
            anchorcolor=blue,
            citecolor=blue
           ]{hyperref}

\makeatletter
\xpatchcmd{\proof}{\@addpunct{.}}{\@addpunct{:}}{}{}
\makeatother

\pgfplotsset{compat=1.18}
\begin{document}

\title{A Novel Hybrid Optical and STAR IRS System for NTN Communications}

\author{Shunyuan Shang,~Emna~Zedini,~\IEEEmembership{Member,~IEEE,}~Abla Kammoun,~\IEEEmembership{Member,~IEEE,}~and~Mohamed-Slim~Alouini,~\IEEEmembership{Fellow,~IEEE} 
\thanks{(\textit{Corresponding author: Shunyuan Shang})}
\thanks{S. Shang, A. Kammoun and M.-S. Alouini are with the Computer, Electrical, and Mathematical Science and Engineering  Division, King Abdullah University of Science and Technology, Thuwal, Makkah Province, Saudi Arabia (e-mail: shunyuan.shang@kaust.edu.sa;abla.kammoun@kaust.edu.sa; slim.alouini@kaust.edu.sa).}
    \thanks{E. Zedini is with the College of Innovation and Technology, University of Michigan-Flint, Flint, MI, USA (e-mail: ezedini@umich.edu).}}

% The paper headers
% \markboth{Journal of \LaTeX\ Class Files,~Vol.~14, No.~8, August~2015}%
% {Shell \MakeLowercase{\textit{et al.}}: FSO IRS }

\maketitle

\begin{abstract}
This paper proposes a novel non-terrestrial networks (NTNs) system that integrates optical intelligent reflecting surfaces (OIRS) and simultaneous transmitting and reflecting Intelligent reflecting surfaces (STAR-IRS) to address critical challenges in next-generation communication networks. The proposed system model features a signal transmitted from the optical ground station (OGS) to the earth station (ES) via an OIRS mounted horizontally on a high altitude platform (HAP). The ES uses an amplify-and-forward (AF) relay with fixed gain for signal relaying, which is then transmitted through a STAR-IRS vertically installed on a building to facilitate communication with both indoor and outdoor users.  The FSO link incorporates (multiple-input multiple-output) MIMO technology, and this paper develops a channel model specifically designed for scenarios where the number of OIRS units exceeds one. For the radio-frequency (RF) link, a novel and highly precise approximation method is introduced, offering superior accuracy compared to traditional approaches based on the central limit theorem (CLT). Closed-form analytical expressions for key performance metrics, including outage probability (OP), ergodic capacity and average bit error rate (BER) are derived in terms of the bivariate Fox-H function for this novel five hops system. Asymptotic expressions at high SNR are also presented, providing insights into system diversity order. 
\end{abstract}


\IEEEpeerreviewmaketitle

\section{Introduction}

Sixth-generation (6G) networks will integrate multi-technology and multi-use-case architectures, encompassing heterogeneous nodes and diverse systems. These networks are poised to support bandwidth-intensive and latency-sensitive applications, including virtual reality (VR), augmented reality (AR), mixed reality (MR), and the broader category of extended reality (XR), which demand terabit-per-second (Tbps) data rates, ultra-low latency, and highly reliable connectivity \cite{dang2020should,amodu2024technical}. Although communication protocols have continuously evolved, terrestrial networks alone remain insufficient to meet the exacting performance requirements of 6G and the expanding global connectivity needs \cite{saad2019vision,amodu2023thz}. Challenges such as coverage gaps, natural disasters leading to service outages, and surges in traffic during major events highlight the necessity of NTNs to offer complementary and dependable services, either on a permanent basis or as circumstances dictate.

Traditionally, NTNs have relied on aerial or flying base stations (BSs). However, this approach faces considerable challenges due to constraints in cost, size, weight, and power (C-SWaP). To address these limitations, intelligent reflecting surfaces (IRS) have been introduced into NTN systems as a promising alternative \cite{amodu2024technical}. IRS technology represents a groundbreaking innovation in wireless communications, comprising multiple small, cost-effective, and energy-efficient devices capable of dynamically controlling and manipulating wireless signals. In doing so, IRS systems significantly enhance communication performance and efficiency \cite{pan2022overview}. 

Compared to flying BSs, IRS-assisted NTNs offer notable advantages in size, weight, cost, and energy efficiency, particularly when employing nearly-passive IRS architectures that eliminate the need for additional RF chains or signal amplification and decoding circuits \cite{amodu2023thz}. There is a growing body of research focusing on IRS-assisted NTNs \cite{shang_enhancing,shaik2024performance,an2024exploiting,tanash2024integrating,guo2023star,baek2023star,malik2022performance,nguyen2022design}. Traditionally, NTNs have relied on aerial or flying base stations (BSs). However, this approach faces considerable challenges due to constraints in cost, size, weight, and power (C-SWaP). To address these limitations, intelligent reflecting surfaces (IRS) have been introduced into NTN systems as a promising alternative \cite{amodu2024technical}. IRS technology represents a groundbreaking innovation in wireless communications, comprising multiple small, cost-effective, and energy-efficient devices capable of dynamically controlling and manipulating wireless signals. In doing so, IRS systems significantly enhance communication performance and efficiency \cite{pan2022overview}.  

{ IRS technologies can be categorized into several types based on their operational principles and functionalities. The most conventional type is the RF-only IRS, which focuses solely on reflecting incident RF signals to improve coverage and mitigate blockage in terrestrial and NTN systems \cite{pan2022overview}. simultaneous transmitting and reflecting IRS (STAR-IRS), supports both reflection and transmission of RF signals within the same unit, thereby enhancing coverage for both indoor and outdoor users in a single deployment \cite{guo2023star}. In contrast, Hybrid IRS (HIRS) also provides reflection and transmission capabilities but is primarily designed for mmWave terrestrial systems, focusing on dynamic beam management, high-directionality links, and precise beam steering in dense urban environments \cite{verma2024efficient}. HIRS often integrates additional RF chains or active elements for complex beamforming, resulting in higher power consumption and system complexity compared to STAR-IRS. While STAR-IRS typically employs a near-passive structure with simpler control mechanisms, making it more energy-efficient and easier to deploy in NTN scenarios, HIRS prioritizes high-precision directional control at the expense of increased Cost, Size, Weight, and Power (C-SWaP) requirements.  In addition, Optical IRS (OIRS) addresses line-of-sight challenges in FSO communications by dynamically redirecting and shaping optical beams to maintain stable high-capacity links under adverse conditions \cite{OIRS1}.}

{
RF and FSO communications each exhibit distinct advantages and limitations, making them appropriate for different operational scenarios. RF links provide robust connectivity, particularly in non-line-of-sight (NLOS) urban environments, but often suffer from limited bandwidth and interference. Conversely, FSO links can deliver extremely high data rates with low latency, yet they are highly dependent on clear line-of-sight (LOS) conditions and are vulnerable to atmospheric effects. Recognizing these complementary characteristics, hybrid RF/FSO systems have been proposed to bridge the connectivity gap between terrestrial RF access networks and high-capacity FSO backbone links, thereby enhancing both coverage and reliability \cite{wang2023uplink,huang2021uplink,qu2022uav}. As envisioned for 6G NTN systems, future communication scenarios will involve extremely heterogeneous environments and highly diverse use cases, ranging from dense urban areas to remote rural regions, and from high-speed vehicular links to low-power IoT applications. These scenarios demand the seamless integration and collaboration of multiple communication technologies to achieve robust, high-capacity, and ultra-reliable connectivity.

Building on these developments, this work proposes a novel NTN architecture that strategically integrates both Optical IRS (OIRS) and STAR-IRS. OIRS technology helps mitigate the LOS dependence of FSO communications by dynamically redirecting and shaping optical signals, thereby maintaining high-capacity links even under challenging atmospheric conditions \cite{sood2018analysis,al2020survey}. Meanwhile, STAR-IRS is designed to overcome obstructions in dense urban environments by simultaneously reflecting and transmitting RF signals, facilitating seamless connectivity for both outdoor and indoor users.
}Among non-terrestrial options like satellites, unmanned aerial vehicles (UAVs), and high altitude platforms (HAPs) stations, HAPs stand out for their cost-effective deployment, avoiding space launch expenses, and their ease of upgrade, repair, and redeployment \cite{abbasi2024haps}. Operating at stratospheric altitudes (17–22 km) with minimal wind and turbulence, HAPs serve as quasi-stationary platforms with applications in broadcasting, internet connectivity, agriculture, environmental monitoring, emergency communication, surveillance, and disaster response \cite{elamassie2023free}. Given these advantages, HAPs are the chosen platform for this work.

{ The proposed integration of multi-unit OIRS and STAR-IRS into a hybrid RF/FSO NTN system finds applications across a diverse range of real-world scenarios. Specifically, this architecture can provide robust emergency communication solutions during natural disasters or network outages, swiftly restoring connectivity in affected urban and rural regions. It is also particularly beneficial in dense urban environments, where obstructions typically challenge traditional communication methods, offering enhanced coverage for smart-city applications, including real-time environmental monitoring, public safety management, and intelligent transportation systems. Furthermore, the integrated system can facilitate reliable high-capacity communication for high-speed transportation networks, supporting continuous connectivity for trains, autonomous vehicles, and aerial mobility services. In remote and underserved rural areas, it effectively bridges the digital divide by delivering stable broadband access, thereby enhancing educational opportunities, telemedicine services, and local economic growth.}

{ The main contributions of this work are as follows:

\begin{itemize}
    \item [--] Development of a novel five-hop NTN architecture that integrates multi-unit OIRS and STAR-IRS. This design addresses the LOS dependency inherent in FSO communications while enabling simultaneous communication with both indoor and outdoor users.

    \item [--] Proposal of an enhanced FSO link model that incorporates multiple OIRS units with MIMO capabilities. This model significantly improves communication stability and system robustness compared to single-unit configurations.

    \item [--] Introduction of a highly accurate approximation method for modeling IRS-assisted RF channels. This approach surpasses the commonly used CLT-based approximations in prior IRS studies, leading to improved modeling precision.

    \item [--] Derivation of closed-form analytical expressions for key performance metrics, including the probability density function (PDF), cumulative distribution function (CDF), and moments of the end-to-end SNR. These results are then used to calculate ergodic capacity, average BER for different modulation schemes, and outage probability (OP), all expressed in terms of the bivariate Fox-H function.

    \item [--] Presentation of accurate asymptotic expressions for OP and average BER in the high signal-to-noise ratio (SNR) regime. These simplified results enable the determination of the system’s diversity order, providing deeper insights into the NTN system’s performance under high-SNR conditions.
\end{itemize}}

{ The remainder of this paper is organized as follows: Section II reviews related work and highlights the research motivation. Section III describes the proposed system architecture and channel models. Section IV presents a statistical analysis of the end-to-end SNR, deriving analytical expressions for various performance metrics along with their asymptotic behavior in the high-SNR regime. Section V discusses numerical and simulation results, and Section VI concludes the paper with final remarks.}

{ \section{Related Work and Motivation}

The integration of IRS into NTNs has garnered significant attention due to their potential in overcoming traditional communication challenges related to coverage, reliability, and resource constraints. Existing studies generally focus on three IRS types: traditional RF-based IRS, STAR-IRS and OIRS.

RF-based IRS have been widely investigated due to their adaptability and ability to dynamically adjust reflection paths, significantly mitigating multipath effects. Shang et al.~\cite{shang_enhancing}, An et al.~\cite{an2024exploiting}, and Tanash et al.~\cite{tanash2024integrating} illustrated how terrestrial IRS enhance communication reliability. Recently, RF-based IRS specifically mounted on HAPs have been explored. Shaik et al.~\cite{shaik2024performance} examined an integrated satellite-HAP-ground communication system, comparing aerial versus terrestrial IRS deployments and demonstrating substantial gains in communication reliability. Benaya et al.~\cite{benaya2024outage} analyzed outage performance for HAP-mounted IRS-assisted multi-user systems, highlighting their effectiveness in providing reliable coverage. Further expanding this, Benaya et al.~\cite{benaya2025outage} investigated THz-enabled IRS-assisted multi-user space-air-ground integrated networks (SAGINs), demonstrating significant performance enhancements in coverage and reliability.

STAR-IRS, which simultaneously transmit and reflect signals, have recently emerged as versatile solutions for complex communication environments. Guo et al.~\cite{guo2023star} and Baek et al.~\cite{baek2023star} demonstrated the versatility of STAR-IRS in urban scenarios, significantly enhancing signal management compared to traditional IRS solutions. Recent research has specifically integrated STAR-IRS with HAP-based systems. Liu et al.~\cite{liu2025star} introduced STAR-IRS-assisted train-to-ground communications within SAGINs, emphasizing improved connectivity and reliability. Wu et al.~\cite{wu2025federated} further explored federated learning applications in STAR-IRS-aided SAGINs, demonstrating advancements in cooperative communication efficiency.

OIRS have been introduced to overcome LOS constraints prevalent in FSO communications. Malik et al.~\cite{malik2022performance} and Nguyen et al.~\cite{nguyen2022design} initially explored single-unit OIRS implementations, highlighting their cost-effectiveness and energy efficiency. Recent efforts specifically combining OIRS with HAP-based systems include the work of Le et al.~\cite{le2024harvested}, who evaluated energy harvesting performance for OIRS-assisted ground-HAP-UAV systems, demonstrating enhanced energy efficiency and communication reliability. Furthermore, Trinh et al.~\cite{trinh2025optical} investigated OIRS applications for improving secret key rates in quantum key distribution (QKD) between HAPs and UAVs, significantly enhancing secure communications. However, existing studies typically employed simplified single-unit OIRS configurations without exploiting the advantages of multiple-input multiple-output (MIMO) techniques.

Despite these valuable contributions, existing literature primarily addresses single IRS technologies independently (either RF or optical). Such isolated approaches neglect the potential synergistic benefits achievable through combined IRS strategies. Moreover, traditional modeling approaches for RF channels predominantly use Central Limit Theorem (CLT)-based approximations, limiting accuracy in practical IRS deployments.

Motivated by these research gaps, this work proposes an advanced NTN architecture that integrates both multi-unit OIRS with MIMO capabilities and STAR-IRS technology. By strategically combining these different IRS types and addressing the limitations of existing modeling approaches, the proposed framework significantly enhances the robustness and overall performance of NTN systems, effectively tackling both the LOS challenges in FSO communications and the complex propagation conditions in RF scenarios.

To mitigate the LOS dependency inherent in FSO systems, OIRS are deployed on the HAP to ensure reliable connectivity. Considering the complexity of urban environments, where obstructions may exist between users and the earth station (ES) that prevent direct communication with users inside buildings, STAR-IRSs are installed on building facades. This arrangement allows STAR-IRSs to overcome obstacles for outdoor users while simultaneously supporting communication with indoor users, thereby ensuring broader coverage. The proposed NTN architecture integrating multi-unit OIRS with MIMO and STAR-IRS has diverse real-world applications: it can restore connectivity in disaster recovery scenarios; enhance urban coverage in smart cities; improve high-speed transportation links; and deliver reliable broadband to remote and rural areas, bridging the digital divide. 

Rather than attempting physical-layer fusion, the proposed architecture achieves hybridization at the system level: OIRS handles the high-throughput HAP–ES FSO backbone, while STAR-IRS ensures wide-area ground coverage in dense urban or obstructed environments. Moreover, although STAR-IRS does not perform signal decoding, it can apply optimized phase shifts to a composite signal (e.g., \(s_1 + s_2\)) on both transmission and reflection paths, effectively steering each component toward its intended user. Assuming channel independence, the residual interference remains negligible compared to the intended signal, thus supporting efficient multi-user access.

Furthermore, unlike prior studies that have largely focused on single-unit OIRS configurations, this work develops a comprehensive channel model for multi-unit OIRS, incorporating MIMO techniques to enhance FSO performance. Finally, a novel and highly accurate approximation method is introduced for modeling the RF channel, which surpasses the CLT-based methods commonly used in previous IRS research \cite{shang_enhancing}. Together, these contributions establish a more robust and efficient NTN system.
}

\section{Channel and System Models}

\begin{figure}[!ht]
\centering\includegraphics[width=0.45\textwidth]{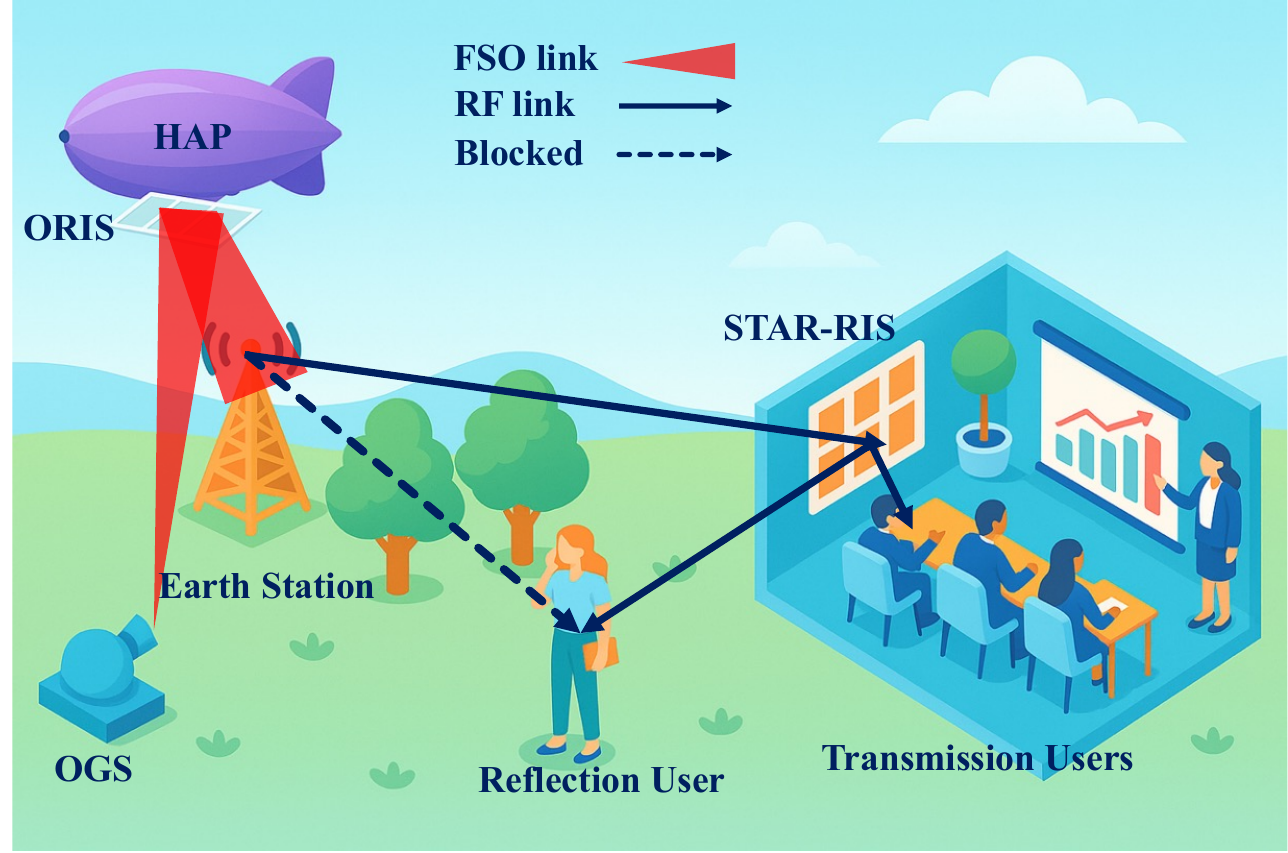}
    \caption{Hybrid OIRS and STAR-IRS System for NTN Communications.}
    \label{figure1}
\end{figure}

As shown in Fig. \ref{figure1}, this paper proposes a novel system that integrates an OIRS and a STAR-IRS. The signal originates from the optical ground station (OGS) and is transmitted to the earth station (ES) via the OIRS, horizontally mounted on the HAP. Given the advantages of amplify-and-forward (AF) relays, including lower computational complexity and reduced latency compared to decode-and-forward (DF) relays, AF relays are ideal for scenarios with limited processing resources or stringent time constraints. Accordingly, the ES employs an advanced AF relay with fixed gain for signal relaying. The relayed signal is then transmitted through the STAR-IRS, which is vertically installed on a building, facilitating communication with users both inside and outside the building.

\subsection{SNR Statistics from OGS to ES (FSO Link)}
For simplicity, we referred to the Users inside the building as “Transmission Users” and those outside the building as “Reflection Users.” In this paper, we used $\ell \in \{R, T\}$ to represent the two different types of Users. 
The heights of the OGS, HAP, ES, STAR-IRS, and Users are denoted by $H_O$, $H_H$, $H_E$, $H_S$, and $H_\ell$, respectively. The distances between these entities are represented as $d_{OH}$ (OGS to HAP), $d_{HE}$ (HAP to ES), $d_{ES}$ (ES to STAR-IRS), and $d_{S\ell}$ (STAR-IRS to Users). The horizontal distances are denoted as $d_{ES0}$ (from the ES to the STAR-IRS) and $d_{S\ell0}$ (from the STAR-IRS to the users). In the FSO link, the zenith angle of the OGS is given by $\zeta_1$, while the zenith angle of the ES is given by $\zeta_2$.

\begin{figure}[!ht]
\centering\includegraphics[width=0.4\textwidth]{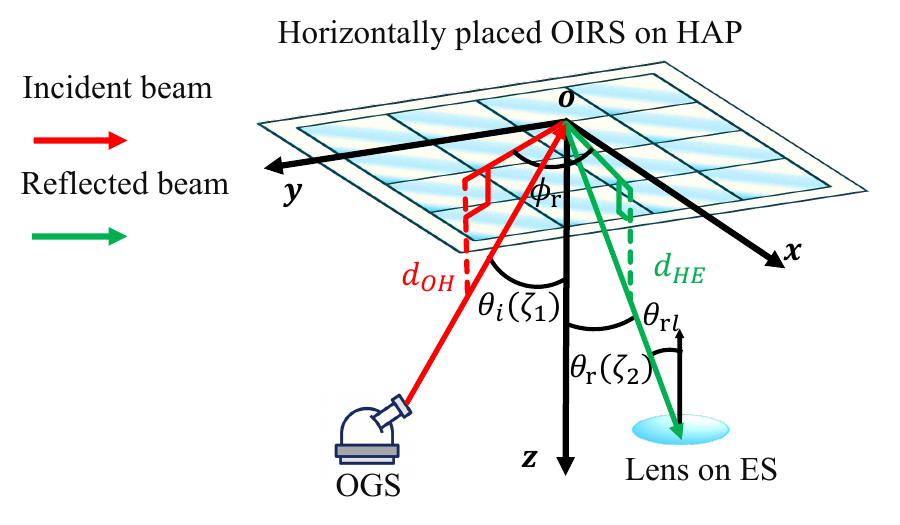}
    \caption{Schematic diagram of the OIRS System.}
    \label{OIRS}
\end{figure}
{ To clearly define the angles involved as the beam passes through the OIRS,  Fig.~\ref{OIRS} illustrates the schematic of the OIRS-assisted transmission system, where the horizontally placed OIRS lies in the \( xoy \) plane and its surface normal is aligned with the \( z \)-axis. The incident beam, emitted from the OGS, arrives at the OIRS with an incidence angle \( \theta_i\), which is defined as the angle between the incoming beam and the surface normal vector. Since the OIRS is horizontally mounted, this incidence angle corresponds to the zenith angle of the OGS, denoted by \( \zeta_1 \). The reflected beam is steered toward the ES, forming a reflection angle \( \theta_r \) with respect to the surface normal. This reflection angle corresponds to the zenith angle of the ES, denoted by \( \zeta_2 \). Unlike conventional specular reflection, the OIRS intelligently adjusts the reflection direction to optimize the received signal power, effectively controlling both \( \theta_r \) and the azimuthal angle \( \phi_r \), which  represents the angle between the projections of the incident and reflected beams onto the \( xoy \) plane. Additionally, \( \theta_{rl} \) denotes the angle between the reflected beam and the optical axis of the receiving lens at the ES, representing the alignment quality of the received signal.}

Let $s(t)$ denote the signal transmitted by the OGS. The signal received by the ES is given by
\begin{align}
    y_{F}(t) = \sqrt{\eta P_{F}} h s(t) + n_{F}(t),
\end{align}
where $P_{F}$ represents the transmit power of the OGS, and $\eta$ denotes the optical-to-electrical conversion factor. The term $n_{F}(t) \sim \mathcal{N}(0, \sigma_{F}^{2})$ models Gaussian noise with a zero mean and variance $\sigma_{F}^{2}$, while $h$ represents the channel gain of the FSO link.
{ In this work, we propose a channel model where the OGS is equipped with \( N_F \) transmit antennas.} The channel gain is determined by three primary factors: attenuation losses ($h_p$), atmospheric turbulence ($h_a$), and geometric and misalignment losses (GML) ($h_g$). Since the propagation distances of multiple optical paths are nearly identical and all optical paths propagate simultaneously, the laser source, the OIRS, and the receiving lens experience consistent jitter across all paths. Consequently, the attenuation losses ($h_p$) and the GML ($h_g$) can be considered consistent across all paths. The overall channel gain is therefore expressed as:
\begin{equation}
    h = h_p h_g h_a = h_p h_g \sum_{i=1}^{N_F}  \tilde{h}_{ai},
\end{equation}
where \( \tilde{h}_{ai}\) represents the atmospheric turbulence effect for the \(i\)-th optical path. 

\subsubsection{Attenuation Loss}
In practical OIRS systems, a portion of the beam's power is absorbed or scattered in addition to being reflected. Let $\zeta_p$ denote the reflection efficiency. For FSO systems operating at a wavelength of \( \lambda = 1550 \, \text{nm} \), typical reflection efficiency values range from 0.7 to 1, as indicated in \cite{zetap1,zetap2}. The absorption at the OIRS can be incorporated into the overall atmospheric loss, represented by $h_{p}$, and is expressed as \cite[Eq.(7)]{OIRS1}
\begin{equation}
    h_{p} =  \zeta_p 10^{-\kappa (d_{OH}+d_{HE})/10},
\end{equation}
where \( \kappa \) is the absorption coefficient. 

\subsubsection{Geometric and misalignment losses}
We adopt the model for geometric and misalignment losses as described in \cite{OIRS1}. This model offers a comprehensive 3D framework that accounts for potential beam misalignment in various directions. By accurately incorporating multiple angular parameters, it effectively represents the signal propagation and reflection process in three-dimensional space. The model enables precise adjustments to the reflected signal, reducing the impact of misalignment. As a result, it is highly suitable for practical applications, ensuring stable signal transmission. The expression for this model is given as \cite[Eq.~(32)]{OIRS1}
\begin{equation}
\label{PDFhg2}
\begin{aligned}
        f_{h_{g}}(h_{g}) = &\frac{\varpi}{A_{0}} \left( \frac{h_{g}}{A_{0}} \right)^{\frac{(1 + q_g^2)\varpi}{2q_g} - 1} \\&\times I_0 \left( -\frac{(1 - q_g^2)\varpi}{2q_g} \ln \left( \frac{h_{g}}{A_{0}} \right) \right),  0 \leq h_g \leq A_{0},
\end{aligned}
\end{equation}
where the function $I_0(\cdot)$ denotes the modified Bessel function of the first kind.
We assume that $\phi_r = \pi$ and $\theta_{rl} = 0$. Under these conditions, the parameters in the formula can be expressed as follows
\[\Omega = \sigma_{u_1}^2 + \sigma_{u_2}^2,
 A_{0} = \text{erf}(\nu_1) \text{erf}(\nu_2), q_g = \sqrt{ \frac{\min\{\sigma_{u_1}^2, \sigma_{u_2}^2\}}{\max\{\sigma_{u_1}^2, \sigma_{u_2}^2\}}},\]
\[
 t_g = \frac{\pi a_l^2}{4 \nu_1 \nu_2} \sqrt{\frac{\pi \, \text{erf}(\nu_1)\text{erf}(\nu_2)}{\nu_1 \nu_2 \exp\left( -(\nu_1^2 + \nu_2^2) \right)}}, \varpi = \frac{(1 + q_g^2)t_g}{4q_g\Omega}, 
\]
\[
\nu_1 =\frac{a_l}{ \omega(d_{OH}\!+\!d_{HE}, \hat\omega_0)}\sqrt{\frac{\pi}{2}},  \nu_2 = \frac{a_l}{ \omega(d_{OH}\!+\!d_{HE}, \omega_0)}\sqrt{\frac{\pi}{2}},
\]
\[
\sigma_{u_1}^2 = \frac{\cos^2 \theta_r}{\cos^2 \theta_i} \sigma_s^2 + \frac{\sin^2(\theta_i + \theta_r)}{\cos^2 \theta_i} \sigma_r^2 + \sigma_l^2,  \sigma_{u_2}^2 = \sigma_s^2 + \sigma_l^2,
\]
where { ${{\omega}}({d}, {\omega}_{0}) = {{\omega}}_{0} \sqrt{1 + \left( \frac{{d}\lambda}{\pi {{\omega}}_{0}^2} \right)^2 }$ is the beam waist for a Gaussian beam with initial beam waist ${{\omega}}_{0}$ and propagation distance ${d}$, $\hat{\omega}_0$ is determined by solving the following equation
\begin{equation}
    \omega(d_{OH},{\hat{\omega}}_{0}) = \frac{\cos(\theta_{r})\omega(d_{OH},{\omega}_0)}{\cos(\theta_{i})}, 
\end{equation}}
$\text{erf}(\cdot)$ represents the error function, $\sigma_s^2$, $\sigma_r^2$, and $\sigma_l^2$ represent the position fluctuations of the laser source, the OIRS, and the receiving lens, respectively. 

\subsubsection{Atmosphere Turbulence}
Given the tens-of-kilometer distance between the HAP and the user, the atmospheric turbulence is modeled using the Gamma-Gamma (GG) distribution, as described in \cite[Eq.~(56), pp.~462]{Laserbeam}:
\begin{align}
\label{pdfha2}
    f_{ {\tilde{h}}_{ai}}( \tilde{h}_{ai})=\frac{2(\tilde{\alpha}\tilde{\beta})^{\frac{\tilde{\alpha}+\tilde{\beta}}{2}}\tilde{h}_{ai}^{\frac{\tilde{\alpha}+\tilde{\beta}}{2}-1}}{\Gamma(\tilde{\alpha}) \Gamma(\tilde{\beta})}  K_{\tilde{\alpha}-\tilde{\beta}}\!\left(2 \sqrt{\tilde{\alpha} \tilde{\beta} \tilde{h}_{ai}}\right)\!,
\end{align}
where $K_a (\cdot)$ indicates the modified Bessel function of the second kind with the order $a$, and $\Gamma (\cdot)$ represents the Gamma function,  $\tilde{\alpha}$   and   $\tilde{\beta}$    are   given   as follows:
\begin{equation}
\tilde{\alpha}={\left[\exp\left(\frac{0.49\sigma_{B}^2}{[1+1.11\sigma_{B}^{12/5} ]^{7/6}}\right)-1\right]}^{-1},
\end{equation}
\begin{align}
    \tilde{\beta}={\left[\exp\left(\frac{0.51\sigma_{B}^2}{[1+0.69\sigma_{B}^{12/5} ]^{5/6}}\right)-1\right]}^{-1},
\end{align} 
where $\sigma_{B}^2$ is expressed as \cite[Eq.~(12)]{ata2022haps}
\begin{align}
       \nonumber  \sigma_{B}^{2}&\,\, =  8.7 {k_w}^{7 / 6}\left(H_H-H_O\right)^{5 / 6} \sec ^{11 / 6}(\zeta_1)\operatorname{Re}\Big(\int_{H_O}^{H_H}   
       \nonumber \\&\times C_{n}^{2}(l)\Big\{\left[\Lambda {\xi_1}^{2}+\mathrm{i} \xi_1(1-\overline{\Theta} \xi_1)\right]^{5 / 6}-\Lambda^{5 / 6} {\xi_1}^{5 / 3}\Big\}\nonumber \\&+ 2.25 {k_w}^{7 / 6}\left(H_H-H_E\right)^{5 / 6} \sec ^{11 / 6}(\zeta_2) \\ 
       \nonumber &\times\operatorname{Re}\Big(\int_{H_E}^{H_H} C_{n}^{2}(l){\left( \frac{l-H_E}{H_H-H_I}\right)}^{5/6} \mathrm{d} l\Big) \nonumber ,
\end{align}
The wave number $k_w = 2\pi / \lambda$ is defined based on the wavelength $\lambda$, measured in meters ($\mathrm{m}$). The secant function is denoted by $\sec(\cdot)$, and the zenith angle for the link between the OGS and the HAP is represented by $\zeta_1$. The Fresnel ratio of the Gaussian beam at the receiver is expressed as $\Lambda = \Lambda_0 / (\Lambda_0^2 + \Theta_0^2)$, where $\Lambda_0 = 2d_{OH} / (k_w W_0^2)$, with $W_0$ representing the beam radius. The beam curvature parameter at the transmitter is defined as $\Theta_0 = 1 - d_{OH} / F_0$. For uplink propagation, the normalized distance parameter is given by $\xi_1 = (l - H_H) / (H_O - H_H)$, while the complementary parameter is expressed as $\overline{\Theta} = 1 - \Theta$, where the beam curvature parameter at the receiver is $\Theta = \Theta_0 / (\Theta_0^2 + \Lambda_0^2)$. Additionally, the turbulence structure constant $C_n^2(l)$ is modeled using the Hufnagel-Valley (HV) profile, as described in \cite[Eq.~(1), pp.~481]{Laserbeam}.
\begin{align}
   \nonumber  C_{n}^{2}(l)&\,\,=0.00594(\omega / 27)^{2}\left(10^{-5} l\right)^{10}\exp (-l / 1000)\\ 
   \nonumber &+2.7 \times 10^{-16} \exp (-l / 1500)+A \exp (-l / 1000),
\end{align}
where $l$ is measured in meters (\(\mathrm{m}\)), $\omega$ represents the root mean square (RMS) wind speed in meters per second (\(\mathrm{m/s}\)), and $A$ denotes the nominal value of $C_{n}^{2}(0)$, as described in \cite[pp.~481]{Laserbeam}.

In \cite{chatzidiamantis2011distribution}, the authors proposed a method to approximate the sum of independent and identically distributed (i.i.d) GG random variables (RV) using a new GG distribution. So the PDF of $f_{h_{a}}$ is given as \cite[Eq. (1)]{chatzidiamantis2011distribution}
\begin{equation}
    f_{h_{a}} = \frac{2(\frac{\alpha \beta}{N_F})^{(\alpha + \beta)/2}}{\Gamma(\alpha) \Gamma(\beta)} h_{a}^{(\alpha + \beta)/2 - 1} K_{\alpha - \beta}\left(2\sqrt{\frac{\alpha \beta}{N_F} h_{a}}\right),
\end{equation}
where $\alpha = N_F \tilde{\alpha} + \varepsilon_{\tilde{\alpha}}$, $\varepsilon_{\tilde{\alpha}} = (N_F - 1) \frac{-0.127 - 0.95\tilde{\alpha} - 0.0058\tilde{\beta}}{1 + 0.00124\tilde{\alpha} + 0.98\tilde{\beta}}$ and $ \beta= N_F \tilde{\beta}$.

{ The SNR of the FSO link \( \gamma_{F} \) can be expressed for both IM/DD and heterodyne detection methods as
\begin{align}
\label{gammaHoverline2}
\gamma_{F}=\overline{\gamma}_{F} h^{r},
\end{align}
where the average SNR is given by \( \overline{\gamma}_{F} = \frac{\left(\eta P_{F}\right)^{r/2}}{\sigma_{F}^{2}} \), where \( r \) depends on the detection method: \( r = 1 \) for heterodyne detection and \( r = 2 \) for intensity modulation/direct detection  (IM/DD). }
The CDF of \( \gamma_{F} \), can be derived using 
an approximation method for both IM/DD and heterodyne detection methods as
{\begin{equation}
\label{CDFH}
\scalebox{1}{$\begin{aligned}
    &F_{\gamma_{F}} (\gamma_{F})= 1 - \frac{\varpi \mathcal{N}_F}{\Gamma(\alpha) \Gamma(\beta_)} \sum_{k=0}^{N_k} \frac{\Gamma(1 + 2k)}{k! \Gamma(1 + k)} \left( \frac{(1 - q_g^2) \varpi}{4q_g} \right)^{2k} 
\\&\times {{\rm G}}_{2k+2, 2k+4}^{2k+4, 0} \left[ \frac{\alpha \beta}{N_FA_{0} h_{p}} \left( \frac{\gamma_{F}}{\overline{\gamma}_{F}} \right)^{\frac{1}{r}} \middle|\!\! 
\begin{array}{c} 
1, {\left\{\frac{(1 + q_g^2) \varpi}{2 q_g} +1\right\}}_{2k+1} \\ 
0, \alpha, \beta, {\left\{\frac{(1 + q_g^2) \varpi}{2 q_g} \right\}}_{2k+1} 
\end{array} 
\!\!\!\right].
\end{aligned}$}
\end{equation}}
The term \( \mathcal{N}_F\) is defined as
\begin{align}
\mathcal{N}_F = \left\{ \sum_{k=0}^{N_k} \frac{2 q_g \Gamma(1 + 2k)}{k! \Gamma(1 + k)(1 + q_g^2)} \left[ \frac{(1 - q_g^2)}{2(1 + q_g^2)} \right]^{2k} \right\}^{-1},
\label{mN_F}
\end{align}
where \( \left\{a\right\}_{2k+1} \) indicates \( 2k+1 \) instances of \( a \). 
{ \begin{proof}
    See Appendix \ref{AP:CDF gammaF}.
\end{proof}}

\subsection{SNR Statistics from ES to Users}
The received signal at the users can be given as
\begin{align}
    y_{\ell}(t)=\sqrt{P_{\ell}}\rho_\ell\sum_{i=1}^{N_R} \Tilde{a}_{i} \Tilde{b}_{\ell,i} s_\ell(t)+n_{R}(t),
\end{align}
{ where \(\rho_\ell\) denotes the power allocation factor for user \(\ell\), the power allocation factors for “Transmission Users” and “Reflection Users”  satisfy the normalization condition $\rho_{\text{T}}^2 + \rho_{\text{R}}^2 = 1$,} where $P_{\ell}$ can be expressed as
\begin{align}
     \nonumber P_{\ell}[dB]& = P_{R}[dB]-L_{\ell}[\mathrm{dB}]+G_{Tx}[dB]+G_{Rx}[dB],
\end{align}
where $L_{\ell}[\mathrm{dB}]=40 \log _{10}\left(d_{\mathrm{H} \mathrm{I}}+d_{\mathrm{I} \mathrm{\ell}}\right)+20 \log _{10}{\left(f\right)}$ is the path loss, $G_{\mathrm{Tx}}$ refers to the transmitter antenna gain (dB) and $G_{\mathrm{Rx}}$ represents the receiver antenna gain (dB). Finally, the SNR for the RF channel, $\gamma_{R,\ell}$, can be formulated as
\begin{equation}
\gamma_{R,\ell}=\bar{\gamma}_{R,\ell}\left|\sum_{i=1}^{N} \Tilde{a}_{i} \Tilde{b}_{\ell,i}\right|^{2}=\bar{\gamma}_{R,\ell}\left|\sum_{i=1}^{N}\tilde{R}_{\ell,n}\right|^{2}=\bar{\gamma}_{R,\ell}R_{\ell}^{2}.
\label{gammalbar}
\end{equation}
where $\overline{\gamma}_{R,\ell}={P_{\ell}}\rho_\ell^2/{{\sigma_R}^{2}}$, $\tilde{a}_i$ and $\tilde{b}_{\ell,i}$ represent the channel coefficients modeled by independent Nakagami-m fading distributions for ES to STAR-IRS and from STAR-IRS to Users channels, respectively. The PDFs of \(\tilde{a}_i\) and \(b_{\ell,i}\) can be expressed as follows
\begin{equation}
\begin{aligned}
f_{a_i}(x) &= \frac{2 \tilde{m}_A^{\tilde{m}_A}}{\Gamma(\tilde{m}_A) \tilde{\Omega}_A^{\tilde{m}_A}} x^{2\tilde{m}_A-1} \exp\left(-\frac{\tilde{m}_A}{\tilde{\Omega}_A}x^2\right), \\
f_{b_{\ell,i}}(x) &= \frac{2 \tilde{m}_\ell^{\tilde{m}_\ell}}{\Gamma(\tilde{m}_\ell) \tilde{\Omega}_\ell^{\tilde{m}_\ell}} x^{2\tilde{m}_\ell-1} \exp\left(-\frac{\tilde{m}_\ell}{\tilde{\Omega}_\ell}x^2\right),
\end{aligned}
\end{equation}
where \(\tilde{m}_A\) and \(\tilde{m}_\ell\) represent the shape parameters for \(\tilde{a}_i\) and \(b_{\ell,i}\), respectively, \(\tilde{\Omega}_A\) and \(\tilde{\Omega}_\ell\) are the corresponding scale parameters, and \(\Gamma(\cdot)\) denotes the Gamma function.

In \cite{tegos2022distribution}, the authors provided the PDF for the product of two Nakagami distributions with different parameters, which can be expressed as \cite[Eq. (2)]{tegos2022distribution}
\begin{equation}
f_{\tilde{R}_{\ell,i}}(x) = \frac{4 \tilde{\Psi}_\ell^{\tilde{m}_A + \tilde{m}_\ell}}{\Gamma(\tilde{m}_A) \Gamma(\tilde{m}_\ell)} x^{\tilde{m}_A + \tilde{m}_\ell - 1} K_{\tilde{m}_A - \tilde{m}_\ell}(2 \tilde{\Psi}_\ell x),
\label{KG}
\end{equation} 
where \(\tilde{\Psi}_\ell = \sqrt{\frac{\tilde{m}_A \tilde{m}_\ell}{\tilde{\Omega}_A \tilde{\Omega}_\ell}}\).

Equation (\ref{KG}) denotes the generalized-K \( (K_G) \) distribution, whose definition appears in \cite[Eq. (1)]{peppas2011accurate}. In the same work \cite{peppas2011accurate}, the authors introduce a highly accurate approximation method for the sum of i.i.d. $K_G$ RVs using a new \( K_G \) distribution.
\begin{equation}
    f_{R_\ell}(x) = \frac{4 \Psi_\ell^{k_\ell + m_\ell}}{\Gamma(m_\ell) \Gamma(k_\ell)} x^{k_\ell + m_\ell - 1} K_{k_\ell - m_\ell}(2 \Psi_\ell x).
    \label{PDFRl}
\end{equation} 

The parameters of the new \( K_G \) distribution can be  determined using the method of moments matching, which are given as 
\begin{equation*}
\scalebox{0.95}{$\left\{
\begin{aligned}
k_\ell &= \left|\frac{-\mathcal{B} + \sqrt{\mathcal{B}^2 - 4\mathcal{A}\mathcal{C}}}{2\mathcal{A}}\right| \\
m_\ell &= \left|\frac{-\mathcal{B} - \sqrt{\mathcal{B}^2 - 4\mathcal{A}\mathcal{C}}}{2\mathcal{A}}\right|
\end{aligned}
\right.
\text{or} 
\left\{
\begin{aligned}
k_\ell &= \left|\frac{-\mathcal{B} - \sqrt{\mathcal{B}^2 - 4\mathcal{A}\mathcal{C}}}{2\mathcal{A}}\right| \\
m_\ell &= \left|\frac{-\mathcal{B} + \sqrt{\mathcal{B}^2 - 4\mathcal{A}\mathcal{C}}}{2\mathcal{A}}\right|
\end{aligned}
\right.,$}
\end{equation*}
where 
\[
\left\{
\begin{aligned}
\mathcal{A} &= \mathbb{E}[R_\ell^6]\mathbb{E}[R_\ell^2] + \left(\mathbb{E}[R_\ell^2]\right)^2 \mathbb{E}[R_\ell^4] - 2\left(\mathbb{E}[R_\ell^4]\right)^2, \\
\mathcal{B} &= \mathbb{E}[R_\ell^6]\mathbb{E}[R_\ell^2] - 4\left(\mathbb{E}[R_\ell^4]\right)^2 + 3\left(\mathbb{E}[R_\ell^2]\right)^2 \mathbb{E}[R_\ell^4], \\
\mathcal{C} &= 2\left(\mathbb{E}[R_\ell^2]\right)^2 \mathbb{E}[R_\ell^4], \\
\mathbb{E}&[R_\ell^{n}] =\! \sum_{j_1=0}^{n} \sum_{j_2=0}^{j_1} \cdots \!\!\!\sum_{j_{(N_R-1)}=0}^{j_{(N_R-2)}} 
    \binom{n}{j_1} \binom{j_1}{j_2} \cdots \binom{j_{(N_R-2)}}{j_{(N_R-1)}} \\&
    \quad \times \mathbb{E}[\tilde{R}_{\ell,1}^{(n-j_1)}] \mathbb{E}[\tilde{R}_{\ell,2}^{(j_1-j_2)}]  \cdots  \mathbb{E}[\tilde{R}_{\ell,N_R}^{(j_{N_R-1})}]\,
\end{aligned}
\right.
\]
where the s-th moments of $\tilde{R}_{\ell,i}$ are given as
\begin{equation}
    \mathbb{E}[\tilde{R}_{\ell,i}^n] = \tilde{\Psi}_\ell^{-n} \frac{\Gamma\left(\tilde{m}_A + \frac{n}{2}\right) \Gamma\left(\tilde{m}_\ell + \frac{n}{2}\right)}{\Gamma(\tilde{m}_A) \Gamma(\tilde{m}_\ell)}.
\end{equation}

The PDF of $\gamma_{R,\ell}$ can be obtained from (\ref{PDFRl}) by applying the random variable transformation in (\ref{gammalbar}) as (\ref{PDFR}),
\begin{equation}
\begin{aligned}
    f_{\gamma_{R,\ell}}(\gamma_{R,\ell}) = \, & \frac{2 \Psi_\ell^{k_\ell + m_\ell}}{\Gamma(m_\ell) \Gamma(k_\ell) \gamma_{R,\ell}}
    \left(\frac{\gamma_{R,\ell}}{\bar{\gamma}_{R,\ell}}\right)^{\frac{k_\ell + m_\ell}{2}}
    \\& \times K_{k_\ell - m_\ell}\left(2 \Psi_\ell \sqrt{\frac{\gamma_{R,\ell}}{\bar{\gamma}_{R,\ell}}}\right).
    \label{PDFR}
\end{aligned}
\end{equation}
By applying \cite[Eq.~(14)]{MeijerGalgorithm} and \cite[p32 Eq. (2.1.5)]{Htran1}, (\ref{PDFR}) can be rewritten as \begin{equation}
    f_{\gamma_{R,\ell}}(\gamma_{R,\ell}) = \frac{1}{\Gamma(m_\ell)\Gamma(k_\ell)\gamma_{R,\ell}} {\rm G}^{2,0}_{0,2} \left[ \frac{\Psi_\ell^2 \gamma_{R,\ell}}{\bar{\gamma}_{R,\ell}} \middle| 
    \begin{array}{c}
     -\\
    k_\ell, m_\ell
    \end{array}
    \right].
    \label{PDFR2}
\end{equation}
Then, by using (\ref{PDFR2}) and applying \cite[Eq.~(26)]{MeijerGalgorithm}, we get the CDF of $\gamma_{R,\ell}$ as (\ref{CDFR}),
\begin{equation}
    F_{\gamma_{R,\ell}}(\gamma_{R,\ell}) = \frac{1}{\Gamma(m_\ell)\Gamma(k_\ell)} {\rm G}^{3,1}_{1,3} \left[ \frac{\Psi_\ell^2 \gamma_{R,\ell}}{\bar{\gamma}_{R,\ell}} \middle| 
    \begin{array}{c}
     1\\
    k_\ell, m_\ell, 0
    \end{array}
    \right].
    \label{CDFR}
\end{equation}

\section{END-TO-END SYSTEM PERFORMANCE}

\subsection{End-to-End SNR Statistics}
The end-to-end SNR for the fixed-gain relaying scheme, assuming negligible saturation effects, can be derived using the expression provided in \cite[Eq.~(28)]{fixedgain} as:
\begin{align}
\label{TSNR}
\gamma_{\ell}=\frac{\gamma_{F} \gamma_{R,\ell}}{\gamma_{R,\ell}+C},
\end{align}
where $C$ is a constant depending on the fixed relay gain.

The CDF and PDF of the overall SNR can be expressed in terms of the bivariate Fox-H function, also known as the Fox-H function of two variables, which are provided in  (\ref{PDFE2E}) and (\ref{CDFE2E}). The implementation of this function is detailed in \cite{Peppas}.
\begin{figure*}
\begin{equation}
\scalebox{1}{$\begin{aligned}
f_{\gamma_{\ell}}(\gamma_{\ell}) =\,&  
\frac{\varpi \mathcal{N}_F\sum_{k=0}^{K_F} \frac{\Gamma(1+2k)}{k! \Gamma(1+k)} 
\left[ \frac{(1-q_g^2)\varpi}{4q_g} \right]^{2k} }{\Gamma(\alpha) \Gamma(\beta) \Gamma(m_\ell) \Gamma(k_\ell)\gamma_{\ell}} \\&\times
{\rm H}^{0,1;3,0;0,2k+3}_{1,0;0,3;2k+3,2k+2}\!\left[\!\!\!
\begin{array}{c}
\left( 1; 1, 1 \right) \\
-\\- \\
(0,1)(k_\ell,1)(m_\ell,1) \\
(1-{\alpha},r)(1-{\beta},r)\left\{\left( 1-\frac{(1+q_g^2)\varpi}{2q_g},r \right)\right\}_{2k+1} \\(1,1)\left\{ \left(-(1+q_g^2)\varpi/(2q_g),r \right)\right\}_{2k+1}
\end{array}\!
\middle\vert\!
\begin{array}{c}
\frac{\Psi_\ell^2 C}{\bar{\gamma}_{R,\ell}}\, , \\\\
\left( \frac{A_0 h_p {N}_F}{\alpha\beta} \right)^r \frac{\bar{\gamma}_H}{\gamma_{\ell}}\!\!
\end{array}
\right],
\end{aligned}$}
\label{PDFE2E}
\end{equation}
\begin{equation}
\scalebox{1}{$\begin{aligned}
F_{\gamma_{\ell}}(\gamma_{\ell}) =\,& 1 - 
\frac{\varpi \mathcal{N}_F\sum_{k=0}^{K_F} \frac{\Gamma(1+2k)}{k! \Gamma(1+k)} 
\left[ \frac{(1-q_g^2)\varpi}{4q_g} \right]^{2k} }{\Gamma(\alpha) \Gamma(\beta) \Gamma(m_\ell) \Gamma(k_\ell)} \\&\times
{\rm H}^{0,1;3,0;0,2k+3}_{1,0;0,3;2k+3,2k+2}\!\left[\!\!\!
\begin{array}{c}
\left( 1; 1, 1 \right) \\
-\\- \\
(0,1)(k_\ell,1)(m_\ell,1) \\
(1-{\alpha},r)(1-{\beta},r)\left\{\left( 1-\frac{(1+q_g^2)\varpi}{2q_g},r \right)\right\}_{2k+1} \\(0,1)\left\{ \left(-(1+q_g^2)\varpi/(2q_g),r \right)\right\}_{2k+1}
\end{array}
\middle\vert
\begin{array}{c}
\frac{\Psi_\ell^2 C}{\bar{\gamma}_{R,\ell}}\,, \\\\
\left( \frac{A_0 h_p {N}_F}{\alpha\beta} \right)^r \frac{\bar{\gamma}_H}{\gamma_{\ell}}\!\!
\end{array}
\right].
\end{aligned}$}
\label{CDFE2E}
\end{equation}
\hrule
\end{figure*}

\begin{proof}
See Appendix \ref{ACDFE2E}.
\end{proof}

\subsection{Performance Metrics}
\subsubsection{Outage Probability}
The OP is defined as the probability that the end-to-end SNR falls below a predetermined threshold, $\gamma_{\rm th}$. By substituting $\gamma_\ell$ with $\gamma_{\rm th}$ in (\ref{CDFE2E}), a unified expression for the OP under both detection methods can be directly obtained.

In (\ref{CDFE2E}), the CDF is expressed using the bivariate Fox-H function, which is intricate and not readily available in widely used mathematical software such as MATLAB or MATHEMATICA. { To provide practical insights and facilitate efficient performance evaluation, we derive the asymptotic expression of the CDF in the high SNR regime. The term asymptotic result refers to an approximate analytical expression that captures the dominant behavior of the CDF as the SNR becomes sufficiently large, while neglecting terms that decay more rapidly and contribute less at high SNR levels. This approach significantly simplifies the CDF, yielding a closed-form expression involving only elementary and standard functions that are readily supported by numerical tools such as MATLAB and MATHEMATICA. Furthermore, it enables the derivation of the system diversity order, which is an essential parameter quantifying the slope of the outage probability curve in the high SNR region and directly reflecting the robustness of the communication system against fading and other channel impairments. Therefore, the asymptotic result serves as both a practical tool for numerical evaluation and a fundamental tool for assessing the system reliability under high SNR conditions. 
}
\begin{table*}[b]
\caption{MODULATION PARAMETERS}
\centering
{ \begin{tabular}{|l|l|l|l|l|l|}
\hline \text { Modulation } & $\boldsymbol{\delta_B}$ & $\boldsymbol{p_B}$ & $\boldsymbol{q}_{\boldsymbol{Bm}}$ & $\boldsymbol{N_B}$ & \text { Detection } \\
\hline \text { M-PSK } & $\frac{2}{\max \left(\log _{2} M, 2\right)}$ & 1 / 2 & $\sin ^{2}\left(\frac{(2 k-1) \pi}{M}\right)\log _{2} M$ & $\max \left(\frac{M}{4}, 1\right)$ & \text { Heterodyne } \\
\hline \text { M-QAM } & $\frac{4}{\log _{2} M}\left(1-\frac{1}{\sqrt{M}}\right)$ & 1 / 2 &$\frac{3(2 k-1)^{2}}{2(M-1)}\log _{2} M $& $\frac{\sqrt{M}}{2}$ & \text { Heterodyne } \\
\hline \text { OOK } & 1 & 1 / 2 & 1 / 2 & 1 & \text { IM/DD } \\
\hline
\end{tabular}}
\label{tab:my_label}
\end{table*}

This results in a simplified CDF expression, as shown in (\ref{CDFE2EA}), which involves only elementary and standard functions that are natively supported by MATLAB and MATHEMATICA. The closer $q_g$ is to 1, the higher the accuracy of this formula.
\begin{proof}
See Appendix \ref{ACDFE2EA}.
\end{proof}
\begin{figure*}
\begin{equation}
\scalebox{1}{$\begin{aligned}
F_{\gamma_\ell}(\gamma_\ell) \underset{\bar{\gamma}_H \gg 1}{\mathop{\approx }}&\sum_{k=0}^{K_F}\frac{ \frac{\Gamma(1+2k)}{k! \Gamma(1+k)} \left( \frac{(1-q_g^2)\varpi}{4q_g} \right)^{2k} }{\left[ \frac{(1+q_g^2)\varpi}{2q_g}-\alpha \right]^{2k+1}} 
\frac{\varpi \mathcal{N}_F \Gamma(\beta - \alpha){\rm G}^{3,1}_{1,3} \left[
\frac{\Psi_\ell^2 C}{\bar{\gamma}_{R,\ell}}
\middle\vert 
\begin{array}{c}
1+\frac{\alpha}{r} \\
0, k_\ell, m_\ell
\end{array}
\right]}
{\alpha \Gamma(\alpha) \Gamma(\beta) \Gamma(m_\ell) \Gamma(k_\ell) \Gamma\left(-\frac{\alpha}{r}\right)} 
\left[ \frac{\alpha \beta}{A_0 h_p {N}_F}\left( \frac{\gamma_\ell}{\bar{\gamma}_H} \right)^{\frac{1}{r}} \right]^{\alpha} 
\\&+ \sum_{k=0}^{K_F} \frac{\frac{\Gamma(1+2k)}{k! \Gamma(1+k)} 
\left( \frac{(1-q_g^2)\varpi}{4q_g} \right)^{2k} }{\left[ \frac{(1+q_g^2)\varpi}{2q_g} -\beta\right]^{2k+1}}\frac{\varpi \mathcal{N}_F \Gamma(\alpha - \beta){\rm G}^{3,1}_{1,3} \left[
\frac{\Psi_\ell^2 C}{\bar{\gamma}_{R,\ell}}
\middle\vert 
\begin{array}{c}
1+\frac{\beta}{r} \\
0, k_\ell, m_\ell
\end{array}
\right]
 }
{\beta \Gamma(\alpha) \Gamma(\beta) \Gamma(m_\ell) \Gamma(k_\ell) \Gamma\left(-\frac{\beta}{r}\right)} 
\left[ \frac{\alpha \beta}{A_0 h_p {N}_F}\left( \frac{\gamma_\ell}{\bar{\gamma}_H} \right)^{\frac{1}{r}} \right]^{\beta} \\&
 + \frac{ 2q_g \mathcal{N}_F \Gamma\left({\alpha} - \frac{(1+q_g^2)\varpi}{2q_g}\right) 
\Gamma\left(\beta - \frac{(1+q_g^2)\varpi}{2q_g}\right){\rm G}^{3,1}_{1,3} \left[
\frac{\Psi_\ell^2 C}{\bar{\gamma}_{R,\ell}} 
\middle| 
\begin{array}{c}
1 + \frac{(1+q_g^2)\varpi}{2r q_g} \\
0, k_\ell, m_\ell
\end{array}
\right]}
{\Gamma(\alpha) \Gamma(\beta) \Gamma(m_\ell) \Gamma(k_\ell) (1+q_g^2)\Gamma\left(-\frac{(1+q_g^2)\varpi}{2r q_g}\right) } 
\left[ \frac{\alpha \beta}{A_0 h_p {N}_F}\left( \frac{\gamma_\ell}{\bar{\gamma}_H} \right)^{\frac{1}{r}} \right]^{\frac{(1+q_g^2)\varpi}{2r q_g}} .
\end{aligned}$}
\label{CDFE2EA}
\end{equation}
\hrule
\end{figure*}

This asymptotic expression is particularly useful for analyzing the system’s diversity order. Based on this expression, the diversity gain of the proposed system can be obtained as:
\begin{align}
\label{diversity}
\mathcal{G}_{d} = \min \left(\frac{\alpha}{r} ,\frac{\beta}{r}, \frac{(1 + q_g^2) \varpi}{2 q_g r} \right).
\end{align}

\subsubsection{Average Bit-Error Rate}
A concise and unified expression for the average BER applicable to various coherent M-ary quadrature amplitude modulation (M-QAM) and M-ary phase shift keying (M-PSK) schemes, as well as the IM/DD on-off keying (OOK) technique, can be expressed as \cite[Eq.~(22)]{dualhopFSO}
\begin{align}
\overline{P}_{e}=&\,\delta_B \sum_{m=1}^{N_{B}} \int_{0}^{\infty} \frac{q_{Bm}^{p_{B}}}{2 \Gamma\left(p_{B}\right)}\gamma_\ell^{p_{B}-1} \exp \left(-q_{B m} \gamma_\ell\right) F_{\gamma_\ell}(\gamma_\ell) d \gamma_\ell,
\end{align}
where $N_{B}$, $\delta_{B}$ , $p_{B}$, and $q_{B m}$ are detailed  as Table. \ref{tab:my_label}  \cite{shang_enhancing}. Define  $I_{\ell,m}$ as
\begin{equation}
\begin{aligned}
I_{\ell,m}=& \int_{0}^{\infty}\frac{q_{Bm}^{p_{B}}}{2 \Gamma\left(p_{B}\right)} \gamma_\ell^{p_{B}-1} \exp \left(-q_{B m} \gamma_\ell\right) F_{\gamma_\ell}(\gamma_\ell) d \gamma_\ell,
\end{aligned}
\label{DefI}
\end{equation}
which is given as \eqref{IE2E} for our end to end link from OGS to Users.
\begin{figure*}
\begin{equation}
\scalebox{1}{$\begin{aligned}
I_{\ell,m} = \,&\frac{1}{2} - 
\frac{\varpi \mathcal{N}_F\sum_{k=0}^{K_F} \frac{\Gamma(1+2k)}{k! \Gamma(1+k)} 
\left( \frac{(1-q_g^2)\varpi}{4q_g} \right)^{2k} }{2\Gamma(p_B)\Gamma(\alpha) \Gamma(\beta) \Gamma(m_\ell) \Gamma(k_\ell)}\\&\times{\rm H}^{0,1;3,0;1,2k+3}_{1,0;0,3;2k+3,2k+3}\!\left[\!\!\!
\begin{array}{c}
\left( 1; 1, 1 \right) \\
-\\- \\
(0,1)(k_\ell,1)(m_\ell,1) \\
(1- {\alpha},r)(1- {\beta},r)\left\{\left( 1-\frac{(1+q_g^2)\varpi}{2q_g},r \right)\right\}_{2k+1} \\(q_{Bm},1)(0,1)\left\{\left( -(1+q_g^2)\varpi/(2q_g),r\right) \right\}_{2k+1}
\end{array}\!\!\!
\middle\vert\!\!\!
\begin{array}{c}
\frac{\Psi_\ell^2 C}{\bar{\gamma}_{R,\ell}}\,, \\\\
\left( \frac{A_0 h_p {N}_F}{\alpha\beta} \right)^r q_{Bm}{\bar{\gamma}_H}\!\!\!\!\!\!
\end{array}
\right],
\end{aligned}$}
\label{IE2E}
\end{equation}
\hrule
\end{figure*}
The average BER for OOK, M-QAM, and M-PSK modulations can then be expressed based on  \eqref{IE2E} as follows
\begin{align}\label{BERExp}
    \overline{{P}}_{e}=\delta_{B} \sum_{k=1}^{N_{B}} I\left(p_{B}, q_{B k}\right).
\end{align}
\begin{proof}
See Appendix \ref{AIE2E}.
\end{proof}
\noindent By substituting (\ref{CDFE2EA}) into (\ref{DefI}) and applying the definition of the Gamma function \(\Gamma(z)\), as outlined in \cite{intetable}, a highly accurate asymptotic expression for the average BER in (\ref{IE2E}) can be derived for large \(\bar{\gamma}_H\) values. The resulting expression is presented in (\ref{IE2EA}).
\begin{figure*}
\begin{equation}
\begin{aligned}
I_{m,\ell}&\underset{\bar{\gamma}_H \gg 1}{\mathop{\approx }} 
\frac{\Gamma\left(p_B - \frac{{\alpha}}{r}\right)  \varpi \mathcal{N}_F \Gamma({\beta} - {\alpha})\sum_{k=0}^{K_F} \frac{\Gamma(1+2k)}{k! \Gamma(1+k)} 
\left(\frac{(1-q_g^2)\varpi}{4q_g}\right)^{2k} 
\left(\frac{(1+q_g^2)\varpi}{2q_g}-{\alpha}\right) ^{-2k-1}}
{2 \Gamma(p_B) {\alpha} \Gamma({\alpha}) \Gamma({\beta}) \Gamma(m_\ell) \Gamma(k_\ell) \Gamma\left(-\frac{{\alpha}}{r}\right)}{\rm G}^{3,1}_{1,3} \left[
\frac{\Psi_\ell^2 C}{\bar{\gamma}_{R,\ell}} 
\middle| 
\begin{array}{c}
1+\frac{{\alpha}}{r} \\
0, k_\ell, m_\ell
\end{array}
\right] \\&\times 
\left[\frac{{\alpha} {\beta}\left(q_{Bm}\bar{\gamma}_H\right)^{-\frac{1}{r}}}{A_0 h_p{N}_F}\right]^{{\alpha}}+\frac{\Gamma\left(p_B - \frac{{\beta}}{r}\right)  \varpi \mathcal{N}_F \Gamma({\alpha} - {\beta}) \sum_{k=0}^{K_F} \frac{\Gamma(1+2k)}{k! \Gamma(1+k)} 
\left(\frac{(1-q_g^2)\varpi}{4q_g}\right)^{2k} 
\left(\frac{(1+q_g^2)\varpi}{2q_g}-{\beta}\right) ^{-2k-1}}
{2 \Gamma(p_B) {\beta} \Gamma({\alpha}) \Gamma({\beta}) \Gamma(m_\ell) \Gamma(k_\ell) \Gamma\left(-\frac{{\beta}}{r}\right)}\\&\times {\rm G}^{3,1}_{1,3} \left[
\frac{\Psi_\ell^2 C}{\bar{\gamma}_{R,\ell}} 
\middle| 
\begin{array}{c}
1+\frac{{\beta}}{r} \\
0, k_\ell, m_\ell
\end{array}
\right]
\left[\frac{{\alpha} {\beta}\left(q_{Bm}\bar{\gamma}_H\right)^{-\frac{1}{r}}}{A_0 h_p {N}_F}\right]^{{\beta}}+ \frac{q_g \mathcal{N}_F \Gamma\left( {\alpha} - \frac{(1+q_g^2)\varpi}{2q_g}\right) 
\Gamma\left( {\beta} - \frac{(1+q_g^2)\varpi}{2q_g}\right) 
\Gamma\left(p_B - \frac{(1+q_g^2)\varpi}{2r q_g}\right)}
{(1+q_g^2) \Gamma(p_B) \Gamma( {\alpha}) \Gamma( {\beta}) \Gamma(m_\ell) \Gamma(k_\ell) \Gamma\left(-\frac{(1+q_g^2)\varpi}{2r q_g}\right)} \\&\times {\rm  G}^{3,1}_{1,3} \left[
\frac{\Psi_\ell^2 C}{\bar{\gamma}_{R,\ell}} 
\middle| 
\begin{array}{c}
1 + \frac{(1+q_g^2)\varpi}{2r q_g} \\
0, k_\ell, m_\ell
\end{array}
\right] \left[\frac{{\alpha} {\beta}\left(q_{Bm}\bar{\gamma}_H\right)^{-\frac{1}{r}}}{A_0 h_p {N}_F}\right]^{\frac{(1+q_g^2)\varpi}{2q_g}} .
\end{aligned}
\label{IE2EA}
\end{equation} 
\hrule
\end{figure*}
\subsubsection{Moments}
The $s$-th moments of $\gamma_{\ell}$ can be demonstrated as
\begin{equation}
\begin{aligned}
&\mathbb{E}\left(\gamma^s_{\ell}\right)= \frac{\varpi \mathcal{N}_F\sum_{k=0}^{K_F} \frac{\Gamma(1+2k)}{k! \, \Gamma(1+k)} 
\left( \frac{(1 - q_g^2) \varpi}{4q_g} \right)^{2k}}{\Gamma({\alpha}) \Gamma({\beta}) \Gamma(m_\ell) \Gamma(k_\ell)}  \\
&\times\frac{\Gamma({\alpha}  + rs) \Gamma({\beta}  + rs)}{ \Gamma(s)} \left[\frac{(1+q_g^2) \varpi }{ 2q_g }+ rs \right]^{-2k-1}\\
& \times G_{1,3}^{3,1} \left[ 
\begin{array}{c}
1-s \\ 
0, k_\ell, m_\ell
\end{array} \middle| 
\frac{\Psi_\ell^2 \mathcal{C}}{\bar{\gamma}_\ell}
\right] \left[ \left( \frac{A_0 h_p {N}_F}{{\alpha} {\beta}} \right)^r\bar{\gamma}_H \right]^s .
\end{aligned}
\label{Moments}
\end{equation}
\begin{proof}
See Appendix \ref{AMoments}.
\end{proof}

\subsubsection{Ergodic Capacity}
The ergodic capacity of the end-to-end system where the FSO link is operating under either heterodyne or IM/DD techniques can be formulated as given in \cite[Eq.(26)]{lapidoth}, as follows
\begin{align}\label{DefC}
\overline{C}\triangleq \mathbb{E}[\ln(1+c_0\,\gamma)]=\int_{0}^{\infty}\ln(1+c_0\,\gamma)f_\gamma(\gamma)\,d\gamma,
\end{align}
where $c_0$ is a constant, taking the value $c_0=1$ for the heterodyne technique ($r=1$) and $c_0=e/(2\pi)$ for the IM/DD technique ($r=2$). 

Then, the bivariate Fox-H function can be used to formulate a unified equation for the ergodic capacity  applicable to both heterodyne and IM/DD types of detection is shown in (\ref{capacityE2E}).
\begin{figure*}
\begin{equation}
\scalebox{1}{$\begin{aligned}
&\overline{C} = 
\frac{\varpi \mathcal{N}_F\sum_{k=0}^{K_F} \frac{\Gamma(1+2k)}{k! \Gamma(1+k)} 
\left( \frac{(1-q_g^2)\varpi}{4q_g} \right)^{2k} }{\Gamma(\alpha) \Gamma(\beta) \Gamma(m_\ell) \Gamma(k_\ell)}\\&\times{\rm H}^{0,1;3,0;1,2k+4}_{1,0;0,3;2k+4,2k+4} \left[\!\!\!
\begin{array}{c}
\left( 1; 1, 1 \right) \\
-\\- \\
(0,1)(k_\ell,1)(m_\ell,1) \\(1,r)
(1- {\alpha},r)(1- {\beta},r)\left\{\left( 1-\frac{(1+q_g^2)\varpi}{2q_g},r \right)\right\}_{2k+1} \\(1,1)(0,1)\left\{ \left(-(1+q_g^2)\varpi/(2q_g),r \right)\right\}_{2k+1}
\end{array}\!\!\!
\middle\vert\!\!\!
\begin{array}{c}
\frac{\Psi_\ell^2 C}{\bar{\gamma}_{R,\ell}}, \\\\
\left( \frac{A_0 h_p {N}_F}{\alpha\beta} \right)^r c_{0}{\bar{\gamma}_H}\!\!\!\!\!
\end{array}
\right].
\end{aligned}$}
\label{capacityE2E}
\end{equation}
\hrule
\end{figure*}
\begin{proof}
See Appendix \ref{AcapacityE2E}.
\end{proof}

% \begin{figure*}
% \begin{equation}
% \scalebox{0.9}{$\begin{aligned}
% \overline{C} = 
% \frac{\varpi \mathcal{N}_F\sum_{k=0}^{K_F} \frac{\Gamma(1+2k)}{k! \Gamma(1+k)} 
% \left( \frac{(1-q_g^2)\varpi}{4q_g} \right)^{2k} }{\Gamma(\alpha) \Gamma(\beta) \Gamma(m_\ell) \Gamma(k_\ell)}{\rm H}^{0,1;3,0;1,2k+4}_{1,0;0,3;2k+3,2k+4} \left[\!\!\!
% \begin{array}{c}
% \left( 1; 1, 1 \right) \\
% -\\- \\
% (0,1)(k_\ell,1)(m_\ell,1) \\(1,r)
% (1- {\alpha},r)(1- {\beta},r)\left\{\left( 1-\frac{(1+q_g^2)\varpi}{2q_g},r \right)\right\}_{2k+1} \\(1,1)(0,1)\left\{ -(1+q_g^2)\varpi/(2q_g),r \right\}_{2k+1}
% \end{array}\!\!\!
% \middle\vert\!\!\!
% \begin{array}{c}
% {\Psi_\ell^2 C}/{\bar{\gamma}_{R,\ell}}\,, \\\\
% \left( \frac{A_0 h_p {N}_F}{\alpha\beta} \right)^r c_{0}{\bar{\gamma}_H}\!\!\!\!\!\!
% \end{array}
% \right].
% \end{aligned}$}
% \label{capacityE2E}
% \end{equation}
% \hrule
% \end{figure*}

\section{NUMERICAL ANALYSIS}
In this section, we present the mathematical framework discussed earlier and validate its accuracy using Monte Carlo simulations, based on the system settings detailed in Table \ref{tab2}. Analytical results are derived and compared with the outcomes of the simulations. The comparison demonstrates a close match between the analytical expressions and the simulated data, confirming the reliability and accuracy of the proposed model.
\begin{table}[!h]
\caption{System Parameters}
\begin{tabular}{cccc}
\hline
Parameters          & Values                      & Parameters     & Values      \\ \hline
$H_O$               & 50 m                        & $H_H$          & 18 km    \\
$H_E$               & 50 m                      & $H_S$       & 100 m      \\
$H_T$               & 100 m                        & $H_R$          & 2 m        \\
$d_{ES0}$   & 500 m   &           $d_{ST0}$      & 10 m      \\
$d_{SR0}$           & 500 m                         & $F_0$      & $\infty$        \\
$\omega$                  & 30 m/s                       & $W_0$     & 1 mm      \\
$\lambda$           & 1550 nm                     &     $K_F$    &   5   \\
   $A$             &      $1.7\times 10^{-13}\text{m}^{-2/3}$               &    $a_l$       &  2.5 cm         \\
$\zeta_2=\theta_r$               & $\pi/7$                           &   $\zeta_p$       &  1     \\ $\kappa$ 
       &    $0.43\times 10^{-3}$                 & $ \phi_r  $          & $\pi$      \\\( \theta_{rl}  \)
  &   0        &$C$     &   1\\$\omega(d_{OH}, \omega_0)$
   &  $4a_l$   &   $\tilde{m}_T$         &   2.5       \\$\tilde{m}_A=\tilde{m}_R$
    &      1.5            &  $\tilde{\Omega}_A=\tilde{\Omega}_\ell$     &   1     \\
$\rho_T$               & 0.8                       &    $\rho_R$          & 0.6       \\$\sigma^2_R$
               &  $10^{-10}$                     & $\gamma_{th}$  & 2 dB\\$P_R$& 0 dB      &$f_c$ &  5 Ghz \\$G_t$ &2 dB & $G_r$& 2 dB\\$N_F$&3&$N_R$&16\\$\sigma_s=\sigma_r=\sigma_l$&$0.5a_l$&$\zeta_1=\theta_i$ &$\pi/6$\\ \hline
\end{tabular}\label{tab2}
\end{table}

\begin{figure}[!ht]
\centering\includegraphics[width=0.5\textwidth]{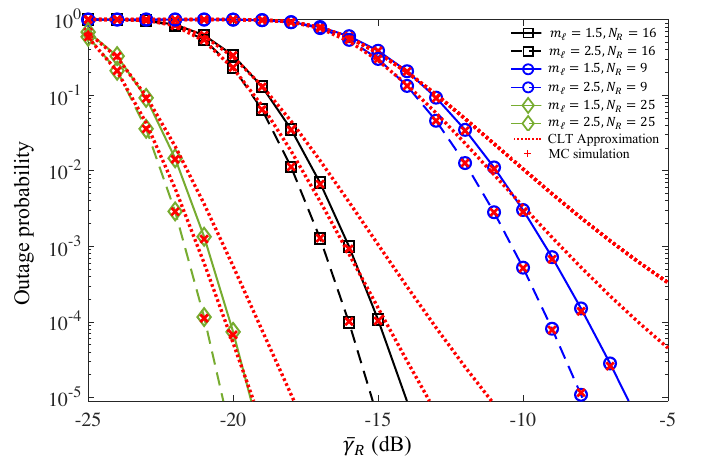}
\caption{Comparison of OP for the RF link under different $m_\ell$ and $N_R$ conditions using approximation methods and MC simulations.}
    \label{figure2}
\end{figure}

First, we analyze the accuracy of the approximation methods used for the RF channel in the paper and compare them with results obtained using the CLT approximation which can be found in \cite[Eq. (30)]{shang_enhancing}. Fig. \ref{figure2} illustrates the OP of the RF link under varying $\tilde{m}_\ell$ and $N_R$ conditions, comparing two different approximation methods and MC simulation results.  As shown in Fig. \ref{figure2}, the approximation methods presented in this paper closely align with the MC simulation results, regardless of whether $\tilde{m}_\ell = 1.5$ or $\tilde{m}_\ell = 2.5$, and irrespective of the values of $N_R = 9$, $N_R = 16$, or $N_R = 25$. In contrast, the approximation based on the CLT only matches the MC simulation results at lower $\bar{\gamma}_R$. As the $\bar{\gamma}_R$ increases, the discrepancy between the CLT-based approximation and the MC results grows significantly. This indicates that the approximation methods used in this paper achieve a high level of accuracy.

\begin{figure}[!ht]
\centering\includegraphics[width=0.5\textwidth]{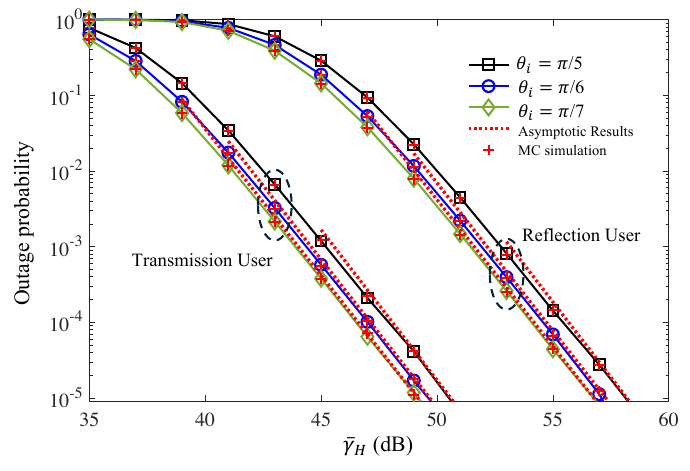}
\caption{Impact of the Incident Angle of the FSO Link, $\theta_i$, on the End-to-End System OP with Heterodyne Detection.}
    \label{figure3}
\end{figure}

After analyzing the approximation accuracy of (\ref{CDFR}) in this study, we proceed to examine the impact of changes in the FSO link on the end-to-end system OP from the OGS to the User. Fig. \ref{figure3} illustrates the effect of varying the incident angle of the FSO link on the end-to-end system, and heterodyne detection is employed. From Fig. \ref{figure3}, it can be observed that as the incident angle increases, the OP of both the Reflection User and Transmission User systems rises consistently. Specifically, for the Reflection User, when $\bar{\gamma}_H = 50$ dB, the OP values under incident angles of $\pi/7$, $\pi/6$, and $\pi/5$ are 0.0079, 0.010, and 0.022, respectively. This increase can be attributed to the growing distance between the OGS and HAP as the incident angle enlarges, leading to greater interference in the FSO link and subsequently higher OP for the overall system. Additionally, Fig. \ref{figure3} highlights that the Transmission User exhibits a lower OP compared to the Reflection User. This is because the distance from the ES  to the Transmission User is shorter, and the Transmission User, being indoors, experiences less interference. Moreover, Fig. \ref{figure3} demonstrates that the closed-form and asymptotic expressions proposed in this paper align well with  MC simulation results, confirming the accuracy of the derived analytical expressions.

\begin{figure}[!ht]
\centering\includegraphics[width=0.5\textwidth]{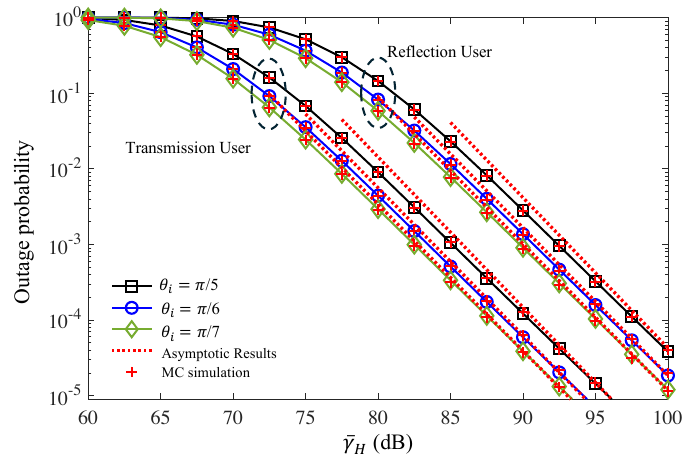}
    \caption{Impact of the Incident Angle, $\theta_i$, of the FSO Link on the End-to-End System OP with IM/DD Detection}
    \label{figure4}
\end{figure}

Fig. \ref{figure4} illustrates the impact of varying the incident angle of the FSO link on the end-to-end system, and IM/DD detection is employed. Fig. \ref{figure4} shows that, overall, the closed-form and asymptotic expressions proposed in this paper align closely with the results of Monte Carlo simulation results, confirming the accuracy of the derived formulas. Compared to heterodyne detection, IM/DD detection results in higher OP, highlighting the advantages of heterodyne detection. Similar to the results with heterodyne detection, the OP increases as the incident angle of the FSO link becomes larger. Specifically, for the Reflection User, when $\bar{\gamma}_H = 50$ dB, the OP values for incident angles of $\pi/7$, $\pi/6$, and $\pi/5$ are 0.058, 0.088, and 0.14, respectively. Compared to the Reflection User, the Transmission User exhibits a lower OP, primarily because of its closer proximity to the emission source and the reduced level of interference it experiences.

\begin{figure}[!ht]
\centering\includegraphics[width=0.5\textwidth]{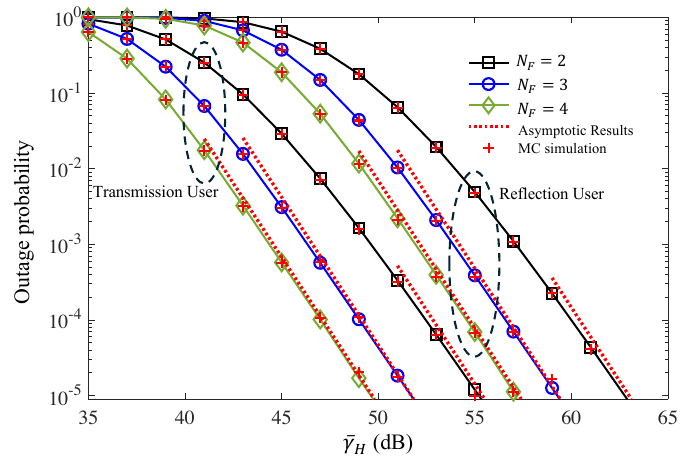}
    \caption{Impact of the Number of Laser Sources $N_F$ on the End-to-End System OP with Heterodyne Detection}
    \label{figure5}
\end{figure}

In addition to the incident angle, the number of laser sources, $N_F$, also significantly impacts the performance of the FSO link. Therefore, analyzing the effect of $N_F$ on the end-to-end system OP from the OGS to the User is meaningful. Fig. \ref{figure5} illustrates the influence of varying $N_F$ on the system's end-to-end OP when heterodyne detection is employed. Under different $N_F$ conditions, the closed-form and asymptotic expressions proposed in this paper align well with the results of Monte Carlo simulation results, validating the accuracy of the derived expressions. As shown in Fig. \ref{figure5}, the end-to-end OP decreases consistently with an increase in $N_F$ for both the Reflection User and Transmission User systems. Specifically, for the Reflection User, when $\bar{\gamma}_H = 50$ dB, the OP values for $N_F = 2$, $N_F = 3$, and $N_F = 4$ are 0.18, 0.043, and 0.0011, respectively. For the same $N_F$, the Transmission User achieves a lower OP than the Reflection User, mainly owing to its shorter distance from the emission source and the reduced interference it encounters.

\begin{figure}[!ht]
\centering\includegraphics[width=0.5\textwidth]{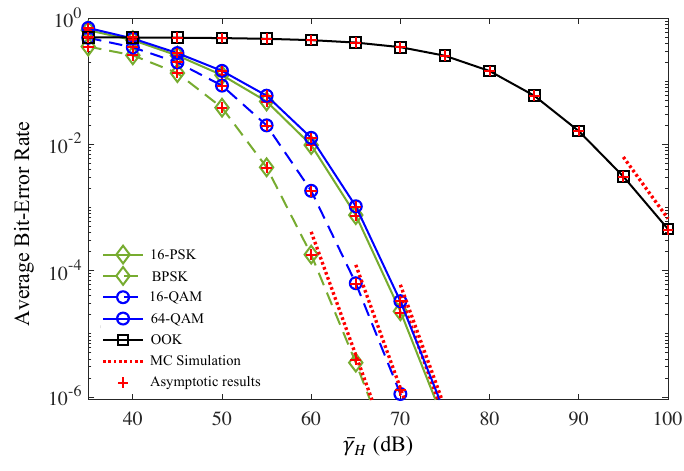}
    \caption{Average BER trends for the Reflection User under various modulation schemes as $\bar{\gamma}_H$ varies}
    \label{figure6}
\end{figure}

In addition to the system's OP, average BER is also a critical performance metric. Therefore, it is necessary to investigate the average BER under different modulation schemes. Fig. \ref{figure6} illustrates the average BER trends for the Reflection User under various modulation schemes as $\bar{\gamma}_H$ varies.  
Overall, the closed-form and asymptotic expressions proposed in this paper align well with the results of Monte Carlo simulation results, confirming the accuracy of the derived expressions. Comparing the results of different modulation schemes, it can be observed that OOK modulation exhibits the highest average BER, primarily due to its use of IM/DD detection. Furthermore, 16-PSK shows a higher average BER compared to B-PSK, and 64-QAM has a higher average BER compared to 16-QAM.

\begin{figure}[!ht]
\centering\includegraphics[width=0.5\textwidth]{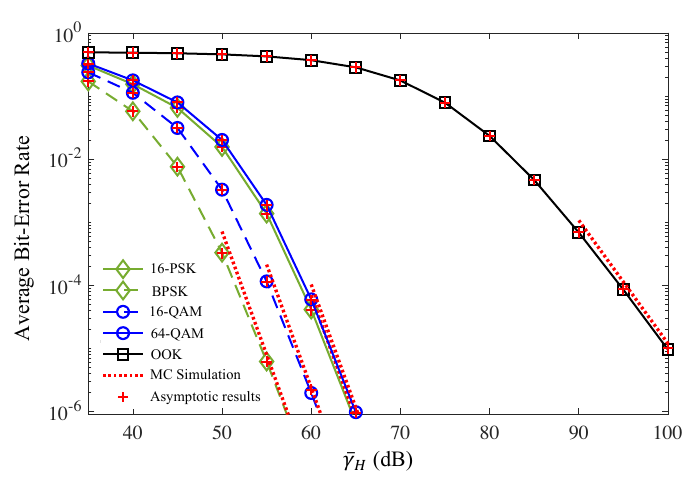}
    \caption{Average BER trends for the Transmission User under various modulation schemes as $\bar{\gamma}_H$ varies}
    \label{figure7}
\end{figure}

Fig. \ref{figure7} illustrates the average BER trends for the Transmission User under various modulation schemes as $\bar{\gamma}_H$ varies. Compared to the Reflection User, the Transmission User achieves a lower average BER when using the same modulation schemes. When comparing different modulation schemes, the Transmission User shows similar results to the Reflection User. 

\begin{figure}[!ht]
\centering\includegraphics[width=0.5\textwidth]{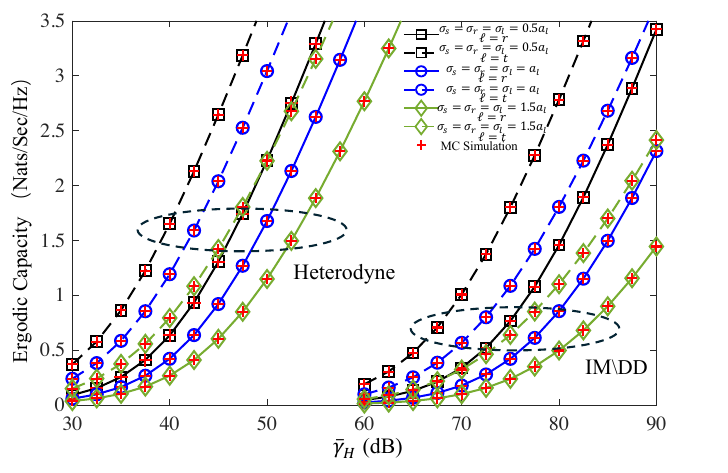}
    \caption{Impact of GML levels in OIRS on channel capacity for two user types under two different detection methods}
    \label{figure8}
\end{figure}

Channel capacity is another critical performance metric in communication systems, making it meaningful to study its variations. Fig. \ref{figure8} illustrates the impact of different levels of GML  in OIRS on the channel capacity for two types of users under two detection methods. As shown in Fig. \ref{figure8}, under identical conditions, the Transmission User achieves higher channel capacity compared to the Reflection User.    Additionally, heterodyne detection provides higher channel capacity than IM/DD detection. Furthermore, with all other conditions remaining constant, an increase in the jitter level of OIRS leads to a reduction in channel capacity. Specifically, for the Transmission User with heterodyne detection, as the jitter level increases, the channel capacities at $\bar{\gamma}_H = 40$ dB are 1.7 \(\mathrm{Nats/s/Hz}\), 1.2 \(\mathrm{Nats/s/Hz}\), and 0.80 \(\mathrm{Nats/s/Hz}\), respectively.
In addition to the FSO link, the RF link also significantly impacts the overall system performance. 

\begin{figure}[!ht]
\centering\includegraphics[width=0.5\textwidth]{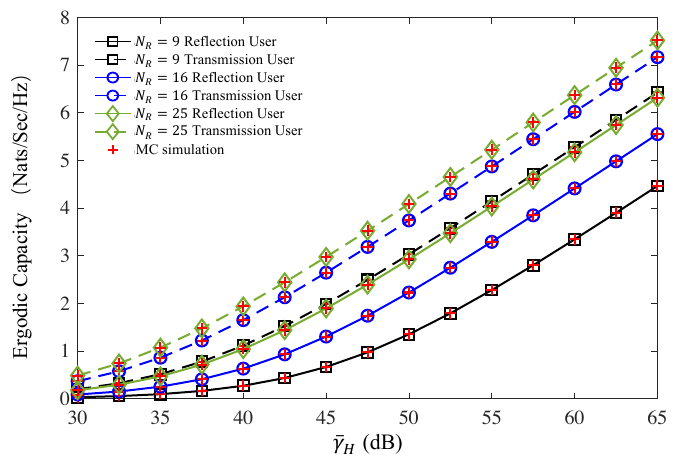}
    \caption{Effect of the number of STAR-IRS elements $N_R$ on channel capacity for two different Users under Heterodyne detection.}
    \label{figure9}
\end{figure}

Fig. \ref{figure9} illustrates the effect of the number of STAR-IRS elements, $N_R$, on the channel capacity for two types of Users under heterodyne detection. As shown in Fig. \ref{figure9}, the system's channel capacity increases consistently with the growth of $N_R$. Specifically, for the Transmission User, when $N_R = 9$, $N_R = 16$, and $N_R = 36$, the channel capacities at $\bar{\gamma}_H = 40$ dB are 1.0 \(\mathrm{Nats/s/Hz}\), 1.6 \(\mathrm{Nats/s/Hz}\), and 1.9 \(\mathrm{Nats/s/Hz}\), respectively.

{
Fig.~\ref{fig:capacity_comparison} illustrates the ergodic capacity versus average SNR \(\bar{\gamma}_H\) for the proposed STAR-IRS configuration in comparison with a conventional IRS-based scheme employing Time Division Multiplexing (TDM). In this benchmark scenario, a traditional reflective IRS is used to alternately serve indoor and outdoor users via orthogonal time slots. While this TDM-based approach allows full power utilization for each user (i.e., \(\rho_\ell = 1\)), it incurs a spectral efficiency penalty by effectively halving the available transmission time for each user, thus reducing the total achievable capacity by a factor of \(1/2\).
\begin{figure}[!ht]
\centering
\includegraphics[width=0.5\textwidth]{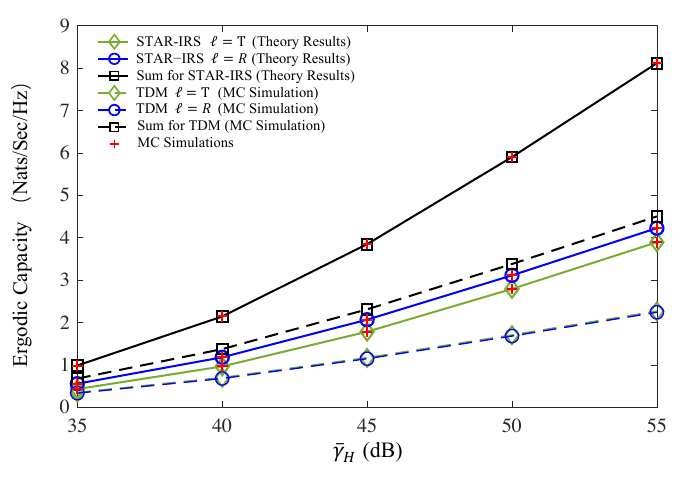}
\caption{Ergodic capacity comparison between the STAR-IRS and TDM-based schemes versus average SNR \(\bar{\gamma}_H\). }
\label{fig:capacity_comparison}
\end{figure}

In contrast, the STAR-IRS configuration enables simultaneous service to both users through its inherent capability to reflect and transmit signals concurrently. Although this introduces a power-splitting effect, i.e., each user receives only a portion of the total transmission power (\(\rho_\ell < 1\)), the two users are served in parallel without time multiplexing. As a result, the total capacity achieved by the STAR-IRS system consistently exceeds that of the TDM-based IRS system across all SNR regimes.

This advantage is clearly depicted in Fig.~\ref{fig:capacity_comparison}, where the solid line (STAR-IRS) lies above the dashed line (conventional IRS with TDM) throughout the entire range of \(\bar{\gamma}_H\). The results demonstrate the superior spectral efficiency and overall system throughput offered by the STAR-IRS architecture in multi-user scenarios.
}

\section{Conclusion}
This paper proposes a novel NTN system that integrates OIRS and STAR-IRS to address critical challenges in next-generation communication networks. The system leverages the strengths of both technologies: the OIRS, mounted on a HAP, mitigates the LOS dependency in FSO communication, while the STAR-IRS, installed on buildings, ensures reliable RF communication with both indoor and outdoor users. Signals are transmitted from the OGS to the ES via the OIRS, and the ES employs an advanced AF relay with fixed gain to efficiently relay signals, reducing latency and computational complexity. The FSO link incorporates MIMO technology, supported by a newly developed channel model tailored for scenarios with multiple OIRS units. For the RF link, a novel and highly accurate approximation method is introduced, surpassing conventional CLT-based approaches. Closed-form expressions for key performance metrics, including ergodic capacity, average BER, and OP are derived using the bivariate Fox-H function for this innovative five hops system. Additionally, asymptotic expressions at high SNR are presented, providing valuable insights into diversity order and high-SNR performance. The accuracy of the proposed closed-form and asymptotic expressions for end-to-end performance metrics, from the OGS to the user, is validated through extensive Monte Carlo (MC) simulations. A comparison with MC results demonstrates that the novel IRS channel approximation method introduced in this paper achieves significantly higher precision than traditional CLT-based methods. Furthermore, simulation results are used to analyze the impact of various parameters of the FSO and RF links on the end-to-end performance metrics, offering critical insights into the behavior of the proposed system. By integrating OIRS and STAR-IRS, this work establishes a robust and efficient framework for NTN communications, advancing hybrid FSO and RF technologies to meet the stringent demands of next-generation networks. { An interesting direction for future work is to incorporate mobility-aware models into the hybrid OIRS/STAR-IRS framework, particularly those capturing Doppler effects, time-varying path-loss, and dynamic beam tracking. This would allow a more comprehensive evaluation of system robustness under realistic user movement scenarios. }

\appendices
\numberwithin{equation}{section}%

{ \section{CDF of the FSO link's SNR }
\label{AP:CDF gammaF}
Using \cite[Eq.~(8.444)]{intetable}, the PDF expression of $h_{g}$ in (\ref{PDFhg2}) can be re-written as
\begin{equation}
\begin{aligned}
    f_{h_{g}}(h_{g}) =&\,\, \frac{\varpi}{A_{0}} \left( \frac{h_{g}}{A_{0}} \right) \left( \frac{(1+q_g^2)\varpi}{2q_g} \right)^{-1} \\&\times\sum_{k=0}^{\infty} \frac{1}{k! \Gamma(1+k)} \left( \frac{(1 - q_g^2)\varpi}{4q_g} \ln \left( \frac{h_{g}}{A_{0}} \right) \right)^{2k}.
    \end{aligned}
\end{equation}
By replacing the infinity with \( N_k \), we can obtain an approximate version of (\ref{PDFhg2}). However, to ensure that the resulting approximation still represents a valid PDF, we need to include a normalization factor \( \mathcal{N}_F \) in \eqref{mN_F}, then we can obtain
\begin{equation}
\begin{aligned}
    f_{h_{g}}(h_{g}) \approx&\,\, \frac{\varpi \mathcal{N}_F}{A_{0}} \left( \frac{h_{g}}{A_{0}} \right) \left( \frac{(1+q_g^2)\varpi}{2q_g} \right)^{-1} \\&\times\sum_{k=0}^{N_k} \frac{1}{k! \Gamma(1+k)} \left( \frac{(1 - q_g^2)\varpi}{4q_g} \ln \left( \frac{h_{g}}{A_{0}} \right) \right)^{2k}.
    \end{aligned}
    \label{ap:A2}
\end{equation}
If (\ref{ap:A2}) is still a valid PDF, then it should satisfy
\begin{equation}
\begin{aligned}
      \sum_{k=0}^{N_k}& \frac{1}{k! \Gamma(1+k)}\int_0^{A_{0}} \left( \frac{(1 - q_g^2)\varpi}{4q_g} \ln \left( \frac{h_{g}}{A_{0}} \right) \right)^{2k}\\& \times\frac{ \mathcal{N}\varpi}{A_{0}} \left( \frac{h_{g}}{A_{0}} \right) \left( \frac{(1+q_g^2)\varpi}{2q_g} \right)^{-1} \mathrm{d}h_{g}=1.
       \end{aligned}
           \label{ap:A3}
\end{equation}
Using the integral formula \cite[Eq.~(4.272.6)]{intetable}  and (\ref{ap:A3}), then the expression of $\mathcal{N}_F$ can be obtained. The PDF of the channel gain $h$ can be written as
\begin{equation}
    f_{h}(h) = \int_{\frac{h}  {A_{0} h_{p}}}^{\infty} f_{h_{g}} \left( \frac{h}{h_{p} h_{a}} \right)  \frac{f_{h_{a}}(h_{a})}{h_{p} h_{a}}\mathrm{d}h_{a}.
    \label{ap:A4}
\end{equation}
By substituting (\ref{PDFhg2}) and (\ref{pdfha2}) into (\ref{ap:A4}), and then applying \cite[Eq.~(14)]{MeijerGalgorithm}, along with the definition of the Meijer-G function from \cite[Eq.~(9.301)]{intetable}, and using \cite[Eq.~(4.272.6)]{intetable}, we can obtain
\begin{equation}
\scalebox{0.9}{$\begin{aligned}
\label{PDFh2}
f_{h}(h)&\,\, = \frac{\varpi \mathcal{N}_F}{h \Gamma(\alpha)\Gamma(\beta)} \sum_{k=0}^{N_k} \frac{\Gamma(1+2k)}{k! \Gamma(1+k)} \left( \frac{(1 - q_g^2) \varpi}{4 q_g} \right)^{2k}\\& \times{\rm G}_{2k+1, 2k+3}^{2k+3,0} \left[ \frac{\alpha \beta h}{N_FA_{0} h_{p}} \middle|
\begin{array}{c}
 {\left\{\frac{(1 + q_g^2) \varpi}{2 q_g} +1\right\}}_{2k+1}\\\alpha, \beta, {\left\{\frac{(1 + q_g^2) \varpi}{2 q_g} \right\}}_{2k+1}
\end{array} 
\right],
\end{aligned}$}
\end{equation}
where \( \left\{a \right\}_{2k+1} \) means there are \( 2k+1 \) instances of \( a \).  The PDF of $\gamma_{F}$ can be obtained from (\ref{PDFh2}) by applying the random variable transformation in (\ref{gammaHoverline2}) as 
{\begin{equation}
\label{PDFH}
\scalebox{1}{$\begin{aligned}
    &f_{\gamma_{F}} (\gamma_{F})=  \frac{\varpi \mathcal{N}_F}{\Gamma(\alpha) \Gamma(\beta_)\gamma_{F}} \sum_{k=0}^{N_k} \frac{\Gamma(1 + 2k)}{k! \Gamma(1 + k)} \left( \frac{(1 - q_g^2) \varpi}{4q_g} \right)^{2k} 
\\&\times {{\rm G}}_{2k+1, 2k+1}^{2k+1, 0} \left[ \frac{\alpha \beta}{N_FA_{0} h_{p}} \left( \frac{\gamma_{F}}{\overline{\gamma}_{F}} \right)^{\frac{1}{r}} \middle|\!\! 
\begin{array}{c} 
 {\left\{\frac{(1 + q_g^2) \varpi}{2 q_g} +1\right\}}_{2k+1} \\ 
 \alpha, \beta, {\left\{\frac{(1 + q_g^2) \varpi}{2 q_g} \right\}}_{2k+1} 
\end{array} 
\!\!\!\right].
\end{aligned}$}
\end{equation}}
Then, by using (\ref{PDFH}) and applying \cite[Eq.~(2.24.2.3)]{prudnikov1}, we obtain the CDF of $\gamma_{F}$ as (\ref{CDFH}).
}

\section{CDF and PDF of the End-to-End SNR}
\label{ACDFE2E}
The CDF of the end-to end SNR $\gamma_{\ell}$ in (\ref{TSNR}) can be formulated as
\begin{align}
\label{A1}
    F_{\gamma_{\ell}}(\gamma_{\ell})=\int_{0}^{\infty} F_{\gamma_{H}}\left(\gamma_{\ell}\left(1+\frac{C}{x}\right)\right) f_{\gamma_{R,{\ell}}}(x)\, dx.
\end{align}
Substituting (\ref{CDFH}) and (\ref{PDFR2}) into (\ref{A1}), yields 
\begin{equation}
\begin{aligned}
&F_{\gamma_\ell}(\gamma_\ell) = 1 - 
\frac{\varpi \mathcal{N}_F\sum_{k=0}^{K_F} \frac{\Gamma(1+2k)}{k! \Gamma(1+k)} 
\left(\frac{(1-q_g^2)\varpi}{4q_g}\right)^{2k}}{\Gamma(\alpha) \Gamma(\beta) \Gamma(m_\ell) \Gamma(k_\ell)} 
\int_{x=0}^\infty \frac{1}{x}\\& \times
{\rm G}^{2k+4,0}_{2k+2,2k+4}\scalebox{0.9}{$ 
\left[\!\frac{\alpha \beta}{A_0 h_p N_F}\!\! \left(\frac{\gamma_\ell}{\bar{\gamma}_H}\right)^{\frac{1}{r}}\!\! \left(1+\frac{C}{x}\right)^{\frac{1}{r}}\!
\middle| \!\!
\begin{array}{c}
1, \left\{\frac{(1+q_g^2)\varpi}{2q_g} + 1\right\}_{2k+1} \\
0, \alpha, \beta, \left\{\frac{(1+q_g^2)\varpi}{2q_g}\right\}_{2k+1}
\end{array}\!\!\!
\right]$}
\\& \times
G^{2,0}_{0,2} 
\left[
\frac{\Psi_\ell^2 x}{\bar{\gamma}_{R,\ell}}
\middle| \begin{array}{cc}
     -  \\k_\ell, m_\ell   
\end{array}
\right] dx.
\end{aligned}
\label{A2}
\end{equation}
Then using the Meijer-G function's primary definition in \cite[Eq.~(9.301)]{intetable}, (\ref{A2}) can be expressed as
\begin{equation}
\label{A3}
\begin{aligned}
&F_{\gamma_\ell}(\gamma_\ell) = 1 - 
\frac{\varpi\mathcal{N}_F\sum_{k=0}^{K_F} \frac{\Gamma(1+2k)}{k! \Gamma(1+k)} 
\left( \frac{(1-q_g^2)\varpi}{4q_g} \right)^{2k} }{\Gamma(\alpha) \Gamma(\beta) \Gamma(m_\ell) \Gamma(k_\ell)} 
\frac{1}{(2\pi i)^2}  \\&\times\!\!\int_{\mathcal{L}_1} \!\!\int_{\mathcal{L}_2}\!\!\!\!
\frac{\Gamma(-s)\Gamma(\alpha-s) \Gamma(\beta-s) }{\Gamma(1-s)} 
\!\!\left[ 
\frac{\Gamma\left(\frac{(1+q_g^2)\varpi}{2q_g} - s\right)}{\Gamma\left(1+\frac{(1+q_g^2)\varpi}{2q_g} - s\right)}
\right]^{1+2k} \\&\times
\Gamma(k_\ell-t) \Gamma(m_\ell-t) 
\left( \frac{\alpha \beta}{A_0 h_p {N}_F} \right)^s 
\left( \frac{\gamma_\ell}{\bar{\gamma}_H} \right)^{\frac{s}{r}} 
\left( \frac{\Psi_\ell^2}{\bar{\gamma}_{R,\ell}} \right)^t 
\\&\times\int_0^\infty x^{t-\frac{s}{r}-1} (x+C)^{\frac{s}{r}} dx \, ds \, dt.
\end{aligned}
\end{equation}
Then, let $s^{\prime}=-rs$ and utilize \cite[Eq.(3.251/11)]{intetable}, (\ref{A3}) can be presented as
\begin{equation}
\scalebox{0.91}{$\begin{aligned}
&F_{\gamma_\ell}(\gamma_\ell) = 1 - 
\frac{\varpi \mathcal{N}_F\sum_{k=0}^{K_F} \frac{\Gamma(1+2k)}{k! \Gamma(1+k)} 
\left( \frac{(1-q_g^2)\varpi}{4q_g} \right)^{2k} }{\Gamma( {\alpha}) \Gamma( {\beta}) \Gamma(m_\ell) \Gamma(k_\ell)} 
\frac{1}{(2\pi i)^2} 
\int_{\mathcal{L}_1} \!\!\int_{\mathcal{L}_2}\\&\times
\frac{\Gamma(t+s^{\prime}) \Gamma(\alpha+rs^{\prime}) \Gamma(\beta+rs^{\prime}) \Gamma\left(\frac{(1+q_g^2)\varpi}{2q_g} + rs^{\prime}\right)}
{\Gamma\left(1+\frac{(1+q_g^2)\varpi}{2q_g} + rs^{\prime}\right)}\!\left( \frac{\bar{\gamma}_H}{\gamma_\ell} \right)^{s^{\prime}} \\&\times
\left( \frac{A_0 h_p {N}_F}{\alpha \beta} \right)^{rs^{\prime}} 
\left[ 
\frac{\Gamma(k_\ell-t) \Gamma(m_\ell-t)}
{\Gamma(1+s^{\prime})} 
\right]^{1+2k}  
\left( \frac{\Psi_\ell^2 C}{\bar{\gamma}_{R,\ell}} \right)^{t} 
ds^{\prime} \, dt.
\end{aligned}$}
\label{A4}
\end{equation}
Using \cite[Eq.~(1.1)]{Mittal}, the desired CDF expression in (\ref{CDFE2E}) can be readily derived.

By differentiating (\ref{A4}) with respect to $\gamma_\ell$, the PDF of $\gamma_\ell$ can be obtained as
\begin{equation}
\scalebox{0.95}{$\begin{aligned}
&f_{\gamma_\ell}(\gamma_\ell) = 
\frac{\varpi \mathcal{N}_F\sum_{k=0}^{K_F} \frac{\Gamma(1+2k)}{k! \Gamma(1+k)} 
\left( \frac{(1-q_g^2)\varpi}{4q_g} \right)^{2k} }{\Gamma( {\alpha}) \Gamma( {\beta}) \Gamma(m_\ell) \Gamma(k_\ell)\gamma_\ell} 
\frac{1}{(2\pi i)^2} 
\int_{\mathcal{L}_1} \!\!\int_{\mathcal{L}_2}\\&\times\frac{\Gamma(t+s^{\prime}) \Gamma(\alpha+rs^{\prime}) \Gamma(\beta+rs^{\prime}) \Gamma\left(\frac{(1+q_g^2)\varpi}{2q_g} + rs^{\prime}\right)}
{\Gamma\left(1+\frac{(1+q_g^2)\varpi}{2q_g} + rs^{\prime}\right)}\!\left( \frac{\bar{\gamma}_H}{\gamma_\ell} \right)^{s^{\prime}} \\&\times
\left( \frac{A_0 h_p {N}_F}{\alpha \beta} \right)^{rs^{\prime}} 
\left[ 
\frac{\Gamma(k_\ell-t) \Gamma(m_\ell-t)}
{\Gamma(s^{\prime})} 
\right]^{1+2k}  
\left( \frac{\Psi_\ell^2 C}{\bar{\gamma}_{R,\ell}} \right)^{t} 
ds^{\prime} \, dt.
\end{aligned}$}
\label{A5}
\end{equation}
Using \cite[Eq.~(1.1)]{Mittal}, the desired PDF expression in (\ref{PDFE2E}) can be readily derived.

\section{Asymptotic Expression for CDF of the End-to-End OP}
For high values of $\overline{\gamma_{H}}$, the following Meijer-G function can be
approximated by using \cite[Eq.~(1.8.4)]{Htran1} as
\label{ACDFE2EA}
\begin{equation}
\begin{aligned}
{\rm G}^{2k+4,0}_{2k+2,2k+4}&\scalebox{0.9}{$ 
\left[\!\frac{\alpha \beta}{A_0 h_p N_F}\!\! \left(\frac{\gamma_\ell}{\bar{\gamma}_H}\right)^{\frac{1}{r}}\!\! \left(1+\frac{C}{x}\right)^{\frac{1}{r}}\!
\middle| \!\!
\begin{array}{c}
1, \left\{\frac{(1+q_g^2)\varpi}{2q_g} + 1\right\}_{2k+1} \\
0, \alpha, \beta, \left\{\frac{(1+q_g^2)\varpi}{2q_g}\right\}_{2k+1}
\end{array}\!\!\!
\right]$}\\&\underset{\bar{\gamma}_H \gg 1}{\mathop{\approx }}\sum_{j}^{4} {\mathcal{H}}_{j}\left[\frac{\alpha \beta}{A_0 h_p N_F}\!\! \left(\frac{\gamma_\ell}{\bar{\gamma}_H}\right)^{\frac{1}{r}}\!\! \left(1+\frac{C}{x}\right)^{\frac{1}{r}}\right]^{h_j},
\end{aligned}
\label{B1}
\end{equation}
where ${h}_j=\left\{0,\alpha,\beta,\frac{(1+q_g^2)\varpi}{2q_g}\right\}$, and
\begin{align}
    \begin{cases}
 \mathcal{H}_{1}=-\Gamma(\alpha) \Gamma(\beta) 
\left[\frac{(1 + q_g^2) \varpi}{2 q_g}\right]^{-2k-1},
\\[1.5ex]
\mathcal{H}_{2}=-\alpha \Gamma(\beta - \alpha) 
\left[\frac{(1 + q_g^2)\varpi}{2q_g} - \alpha\right]^{-2k-1},\\[1.5ex]
\mathcal{H}_{3}=-\beta \Gamma(\alpha - {\beta}) 
\left[\frac{(1 + q_g^2)\varpi}{2q_g} - {\beta}\right]^{-2k-1},
 \\[1.5ex]
\mathcal{H}_{4}=- \frac{(2k+1)2q_g}{(1+q_g^2)\varpi} 
\Gamma\left(\alpha - \frac{(1+q_g^2)\varpi}{2q_g}\right) 
\Gamma\left({\beta} - \frac{(1+q_g^2)\varpi}{2q_g}\right).
\end{cases}
\end{align}
Substituting (\ref{B1}) into (\ref{A2}) and applying the integral in \cite[Eq.(3.194/3)]{intetable}, the asymptotic expression for end to end OP in (\ref{CDFE2EA}) can be obtained.

\section{Average BER}
\label{AIE2E}
Substituting (\ref{A4}) into (\ref{DefI}) gives:
\begin{equation}
\begin{aligned}
&I_{\ell,m} 
= \frac{1}{2} 
- \frac{q_{Bm}^ {p_B}}{2 \Gamma(p_B)} 
\frac{\varpi \mathcal{N}_F\sum_{k=0}^{K_F} \frac{\Gamma(1+2k)}{k! \Gamma(1+k)} 
\left(\frac{(1-q_g^2)\varpi}{4q_g}\right)^{2k} }{\Gamma({\alpha}) \Gamma({\beta}) \Gamma(m_\ell) \Gamma(k_\ell)} 
\frac{1}{(2\pi i)^2}\\&\times 
\int_{\mathcal{L}_1} \!\!\int_{\mathcal{L}_2} \!\!\!
\Gamma\left(t + s\right) \Gamma(k_\ell - t) \Gamma(m_\ell - t) \Gamma(-t) 
\frac{\Gamma({\alpha} + rs) \Gamma({\beta} + rs)}{\Gamma(1+s)} 
\\&\times\left[
\frac{\Gamma\left(\frac{(1+q_g^2)\varpi}{2q_g} + rs\right)}
{\Gamma\left(1 + \frac{(1+q_g^2)\varpi}{2q_g} + rs\right)}
\right]^{1+2k} 
\left(\frac{\Psi_\ell^2 C}{\bar{\gamma}_{R,\ell}}\right)^t 
\left(\frac{A_0 h_p N_F}{{\alpha} {\beta}}\right)^{rs} \\&\times 
\left(\bar{\gamma}_H\right)^s
\int_{0}^\infty \gamma_\ell^{p_B-s-1} \exp(-q_{Bm} \gamma_\ell) d\gamma_\ell \, ds \, dt.
\end{aligned}
\label{C1}
\end{equation}
Using the definition of Gamma Function $\Gamma(z)$ in \cite{intetable}, $\Gamma(z)=\int_0^{\infty} t^{z-1} e^{-t} d t$, (\ref{C1}) can be expressed as
\begin{equation}
\begin{aligned}
&I_{\ell,m}=  \frac{1}{2} 
- \frac{\varpi \mathcal{N}_F\sum_{k=0}^{K_F} \frac{\Gamma(1+2k)}{k! \Gamma(1+k)} 
\left(\frac{(1-q_g^2)\varpi}{4q_g}\right)^{2k} }{2 \Gamma(p_B) \Gamma({\alpha}) \Gamma({\beta}) \Gamma(m_\ell) \Gamma(k_\ell)} 
\frac{1}{(2\pi i)^2} 
\int_{\mathcal{L}_1} \int_{\mathcal{L}_2} \\&\times 
\Gamma\left(t + \frac{s}{r}\right) \Gamma(k_\ell - t) \Gamma(m_\ell - t) \Gamma(-t) 
\frac{\Gamma({\alpha} + rs) \Gamma({\beta} + rs)}{\Gamma(1+s)} 
\\&\times \left[
\frac{\Gamma\left(\frac{(1+q_g^2)\varpi}{2q_g} + rs\right)}
{\Gamma\left(1 + \frac{(1+q_g^2)\varpi}{2q_g} + rs\right)}
\right]^{1+2k} 
\left(\frac{\Psi_\ell^2 C}{\bar{\gamma}_{R,\ell}}\right)^t 
\left(\frac{A_0 h_p N_F}{{\alpha} {\beta}}\right)^{rs} 
\\&\times \left(q_{Bm} \bar{\gamma}_H\right)^s 
\Gamma(p_B - s) \, ds \, dt.
\end{aligned}
\label{C2}
\end{equation}
By applying \cite[Eq.~(1.1)]{Mittal} to (\ref{C2}), the expression in (\ref{IE2E}) can be straightforwardly derived.

\section{Moments}
\label{AMoments}
Using \cite[Eq.~(2.3)]{Mittal}, the PDF of $\gamma_\ell$ in (\ref{PDFE2E}) can be re-written as
\begin{equation}
\scalebox{1}{$
\begin{aligned}
&f_{\gamma_\ell}(\gamma_\ell) = \frac{\varpi \mathcal{N}_F\sum_{k=0}^{K_F} \frac{\Gamma(1+2k)}{k! \, \Gamma(1+k)} }{\gamma_\ell \Gamma( {\alpha}) \Gamma( {\beta}) \Gamma(m_\ell) \Gamma(k_\ell)} 
\left( \frac{(1 - q_g^2)\varpi}{4q_g} \right)^{2k} \int_0^\infty x^{-1}\\
& \times  \exp(-x) {\rm H}_{0,3}^{3,0} 
\left[
\begin{array}{c}-\\
(0,1)(k_\ell,1)(m_\ell,1)
\end{array}
\middle| 
\frac{\Psi_\ell^2 \mathcal{C}x}{\bar{\gamma}_\ell}
\right] {\rm H}_{2k+1,2k+3}^{2k+3,0} \\
&\!\!
\left[\!\!
\begin{array}{c}
 (0,1) \left\{\left( 1+\frac{(1+q_g^2)\varpi}{2q_g},r \right)\right\}_{2k+1}\\({\alpha},r)({\beta},r) \left\{\left( \frac{(1+q_g^2)\varpi}{2q_g},r \right)\right\}_{2k+1}
\end{array}\!\!
\middle| 
\left( \frac{ {\alpha} {\beta}} {A_0 h_p {N}_F}\right)^r 
\frac{\gamma_\ell}{\bar{\gamma}_H x}
\right] dx
\end{aligned}$}
\end{equation}

Utilizing \cite[Eq.~(1.59)]{Htran2} and \cite[Eq.~(2.8)]{Htran2}, the moments of $\gamma_\ell$ can be evaluated according to (\ref{Moments}).

\section{Capacity}
\label{AcapacityE2E}

Substituting (\ref{A5}) into (\ref{DefC}) results in 
\begin{equation}
\begin{aligned}
&\overline{C} = \frac{\varpi \mathcal{N}_F}{\Gamma( {\alpha}) \Gamma( {\beta}) \Gamma(m_\ell) \Gamma(k_\ell)}
\sum_{k=0}^{K_F} \frac{\Gamma(1+2k)}{k! \Gamma(1+k)}
\left(\frac{(1-q_g^2)\varpi}{4q_g}\right)^{2k}\\&\times
\frac{1}{(2\pi i)^2} \int_{\mathcal{L}_1} \int_{\mathcal{L}_2} 
\Gamma\left(t + \frac{s}{r}\right) \Gamma(k_\ell - t) \Gamma(m_\ell - t) \Gamma(-t) 
\\&\times\frac{\Gamma( {\alpha} + rs) \Gamma( {\beta} + rs)}{\Gamma(s)}
\left[
\frac{\Gamma\left(\frac{(1+q_g^2)\varpi}{2q_g} + rs\right)}
{\Gamma\left(1 + \frac{(1+q_g^2)\varpi}{2q_g} + rs\right)}
\right]^{1+2k} \!\!\!\!\!
\left(\frac{\Psi_\ell^2 C}{\bar{\gamma}_\ell}\right)^t \\&\times
\left(\frac{A_0 h_p N_F}{ {\alpha}  {\beta}}\right)^{rs} 
(\bar{\gamma}_H)^s 
\int_{0}^\infty \ln(1 + c_0 \gamma) \gamma^{-1-s} d\gamma \, ds \, dt.
\end{aligned}
\label{D1}
\end{equation}
Using the integral identity $\int_0^{\infty} x^{\mu-1} \ln (1+\gamma x) d x=\frac{\pi}{\mu \gamma^\mu \sin \mu \pi}$ in \cite[Eq.~(4.293/10)]{intetable} along with  $\Gamma(z) \Gamma(1-z)=\frac{\pi}{\sin \pi z}$ in \cite[Eq.(2), p99]{specialfunc}, and applying (1.1) of \cite{Mittal}, the ergodic capacity can be derived in (\ref{capacityE2E}).

\bibliographystyle{IEEEtran}
\bibliography{reference}

\end{document}